\documentclass[aps,showpacs,pre,superscriptaddress,nofootinbib,notitlepage,a4paper]{revtex4}

\usepackage{amsmath}
\usepackage{amsfonts}
\usepackage{amssymb}
\usepackage{xcolor}

\usepackage{graphicx}
\usepackage{psfrag}
\usepackage[naturalnames]{hyperref}

\newcommand{\beq}{\begin{equation}} 
\newcommand{\eeq}{\end{equation}}
\newcommand{\bem}{\begin{multline}}
\newcommand{\bes}{\begin{split}}
\newcommand{\ees}{\end{split}} 
\newcommand{\bea}{\begin{align}}
\newcommand{\eea}{\end{align}}

\newcommand{\us}{\underline{\sigma}}

\newcommand{\ut}{\underline{\tau}}
\newcommand{\ad}{\alpha_{\rm d}}
\newcommand{\asat}{\alpha_{\rm sat}}
\newcommand{\cd}{c_{\rm d}}
\newcommand{\gd}{\gamma_{\rm d}}
\newcommand{\ar}{\alpha_{\rm r}}
\newcommand{\gr}{\gamma_{\rm r}}
\newcommand{\g}{\gamma}
\newcommand{\s}{\sigma}
\newcommand{\dd}{{\rm d}}
\newcommand{\tih}{\widetilde{h}}
\newcommand{\tiu}{\widetilde{u}}
\newcommand{\tC}{\widetilde{C}}

\newcommand{\heta}{\widehat{\eta}}
\newcommand{\hmu}{\widehat{\mu}}
\newcommand{\hnu}{\widehat{\nu}}
\newcommand{\hw}{\widehat{w}}

\newcommand{\hP}{\widehat{P}}
\newcommand{\hQ}{\widehat{Q}}
\newcommand{\hR}{\widehat{R}}
\newcommand{\hm}{\widehat{m}}
\newcommand{\ind}{\mathbb{I}}
\newcommand{\ve}{\varepsilon}
\newcommand{\N}{{\cal N}}
\newcommand{\eqd}{\overset{\rm d}{=}}
\newcommand{\tow}{\overset{\rm w}{\to}}
\newcommand{\di}{\partial i}
\newcommand{\da}{\partial a}
\newcommand{\dima}{\partial i \setminus a}
\newcommand{\dami}{\partial a \setminus i}

\begin{document}

\bibliographystyle{myunsrt}

\title{The asymptotics of the clustering transition for random constraint satisfaction problems}

\author{Louise Budzynski}
\affiliation{Laboratoire de physique de l'Ecole normale sup\'erieure, ENS, Universit\'e PSL, CNRS, Sorbonne Universit\'e, Universit\'e de Paris, F-75005 Paris, France} 

\author{Guilhem Semerjian}
\affiliation{Laboratoire de physique de l'Ecole normale sup\'erieure, ENS, Universit\'e PSL, CNRS, Sorbonne Universit\'e, Universit\'e de Paris, F-75005 Paris, France} 

\begin{abstract}
Random Constraint Satisfaction Problems exhibit several phase transitions when their density of constraints is varied. One of these threshold phenomena, known as the clustering or dynamic transition, corresponds to a transition for an information theoretic problem called tree reconstruction. In this article we study this threshold for two CSPs, namely the bicoloring of $k$-uniform hypergraphs with a density $\alpha$ of constraints, and the $q$-coloring of random graphs with average degree $c$. We show that in the large $k,q$ limit the clustering transition occurs for  $\alpha = \frac{2^{k-1}}{k} (\ln k + \ln \ln k + \gamma_{\rm d} + o(1))$, $c= q (\ln q + \ln \ln q + \gamma_{\rm d}+ o(1))$, where $\gamma_{\rm d}$ is the same constant for both models. We characterize $\gamma_{\rm d}$ via a functional equation, solve the latter numerically to estimate $\gamma_{\rm d} \approx 0.871$, and obtain an analytic lowerbound  $\gamma_{\rm d} \ge 1 + \ln (2 (\sqrt{2}-1)) \approx 0.812$. Our analysis unveils a subtle interplay of the clustering transition with the rigidity (naive reconstruction) threshold that occurs on the same asymptotic scale at $\gamma_{\rm r}=1$.

\end{abstract}

\maketitle

\tableofcontents

\section{Introduction}

A Constraint Satisfaction Problem (CSP) is a system of $N$ variables that can take discrete values, with $M$ constraints that each enforce some requirements on a subset of the variables, a solution of the CSP being an assignement of the variables which satisfies simultaneously all the constraints. The two examples of CSPs that will appear in this paper are the graph coloring and the hypergraph bicoloring problems. In the first one the variables are placed on the vertices of a graph, they have $q$ possible values, to be interpreted as colors, and each edge of the graph enforce the constraint that the two vertices at its ends take different colors. The hypergraph bicoloring problem is similarly defined on an hypergraph, with hyperedges linking subsets of $k$ (instead of two for a graph) vertices; the variables on the vertices can take two colors, and the constraint associated to each hyperedge is that both colors are present among its $k$ adjacent vertices.

CSPs can be studied from several different perspectives; computational complexity theory~\cite{GareyJohnson79,Papadimitriou94} classifies them according to their worst-case difficulty, assessed by the existence or not of an efficient algorithm (running in a time polynomial in $N,M$) able to solve (i.e. to determine the existence or not of a solution) all their possible instances. Another line of work~\cite{MonassonZecchina99b,BiroliMonasson00,MezardParisi02,MertensMezard06,krzakala2007gibbs,AchlioptasRicci06,AchlioptasCoja-Oghlan08,molloy_col_freezing,ding2014proof}, in which this paper finds its place, concentrates on the characterization of the typical instances of a CSP, where typical refers to a random ensemble of instances, most importantly the one obtained by drawing the $M$ constraints uniformly at random. In the examples mentioned above this corresponds to study $G(N,M)$ Erd\H{o}s-R\'enyi random graphs, or their hypergraph generalization. One striking feature of these models is the appearance of phase transitions, or threshold phenomena, when the large size limit (also called thermodynamic limit) $N,M \to \infty$ is taken with a fixed value of the control parameter $\alpha = M/N$, the density of constraints per variable. At these phase transitions some properties that were true with high probability (w.h.p., i.e. with a probability going to one in the large size limit) for a certain value of $\alpha$ become false w.h.p. when $\alpha$ is modified infinitesimally. For instance the satisfiability threshold $\asat$ separates an underconstrained, satisfiable regime $\alpha < \asat$ where typical instances of a random CSP do admit solutions from an overconstrained regime $\alpha > \asat$ where no solution typically exists.

Several other phase transitions occur in the satisfiable phase $\alpha < \asat$, at which the structure of the set of solutions undergoes qualitative changes; in this paper we concentrate on one of them, the so-called clustering (or dynamic) transition that occurs at a critical control parameter denoted $\ad$. This transition can be defined in various ways; the name clustering emphasizes the drastic change of the shape of the set of solutions, viewed as a subset of the whole configuration space. Below $\ad$ the set of solutions of typical instances is rather well-connected, any solution can be reached from any other one by a rearrangement of a non-extensive number of variables. Above $\ad$ the solution set splits in a large number of distinct groups of solutions, called clusters, which are internally well-connected but well-separated one from the other. In the cavity method~\cite{MezardParisi01} treatment of the random CSPs $\ad$ is defined as the appearance of a solution of the one step of Replica Symmetry Breaking (1RSB) equation with Parisi breaking parameter $m=1$, and can also be interpreted as the birth of some long-range correlation between the spin variables under the uniform probability measure over the set of solutions. This correlation is not the usual two-point function but rather a point-to-set correlation function~\cite{MontanariSemerjian06b}, that measures how much information on the value of one spin variable (the point) is provided by the observation of all spins at distance $n$ from it (the set). This correlation decays to zero at large distance for $\alpha < \ad$, and becomes long-ranged for $\alpha > \ad$; this was shown in~\cite{MontanariSemerjian06b} to imply the divergence of the relaxation time of equilibrium dynamics, which justifies the terminology ``dynamic'' and the notation $\ad$.

Erd\H{o}s-R\'enyi random graphs and hypergraphs locally converge, in the thermodynamic limit, to trees and hypertrees: the neighborhood within a fixed distance around an arbitrarily chosen vertex is, with probability one, acyclic. The local limit tree is generated by a Galton-Watson branching process, in which each vertex gives rise to a random number of offsprings with a Poisson distribution of average $c= \alpha k$. As a consequence the computation of the point-to-set correlation function in random graphs can be reduced, within the hypotheses of the cavity method, to an equivalent computation on a tree. The latter was actually studied previously under the name of tree reconstruction problem in an information theoretic perspective~\cite{MoPe03}, see~\cite{MezardMontanari06} for the first discussion of the connection between 1RSB equations and the reconstruction problem. Thanks to the branching structure of trees the generation of uniform configurations, on which the correlation has to be computed, can be done in a recursive, broadcast way, and interpreted as the noisy transmission of information from a root variable to many distant receivers. In this perspective $\ad$ corresponds to the tree reconstruction solvability threshold, the observation of the variables at the $n$-th generation of the tree provides an information on the root that survives the $n\to\infty$ limit if and only if $\alpha > \ad$.

The value of $\ad$ depends of course on the CSP under consideration and of its parameters ($k,q$ in the two examples defined above), and can be estimated numerically by solving a functional equation that arises from the 1RSB formalism or its tree reconstruction interpretation (see for instance~\cite{ZdeborovaKrzakala07} for numerical values in the coloring case, and~\cite{DallAstaRamezanpour08,gabrie2017phase} for the hypergraph bicoloring). There is in general no explicit analytical expression for $\ad$, but bounds on its value~\cite{MoPe03,MezardMontanari06} and asymptotic expansions for large $k,q$~\cite{Sly08,MoReTe11_recclus,SlyZhang16} complement its numerical determination (that becomes very difficult when $k,q$ grow). A relatively simple upperbound on $\ad$ can be obtained by analyzing the so-called naive reconstruction procedure~\cite{Semerjian07,Sly08}, where one asks whether the configuration of the variables at a large distance from a root variable implies unambiguously the value of the latter (instead of merely bringing some information on it). This property undergoes a phase transition at the ``rigidity'' threshold $\ar$, with obviously $\ad \le \ar$, which can be alternatively interpreted as the appearance of ``hard fields'' in the solution of the 1RSB/reconstruction equations, or frozen variables taking the same value in all the solutions of a cluster. The analytic determination of $\ar$ is much simpler than the one of $\ad$, because it corresponds to the bifurcation of a scalar (instead of functional) fixed-point equation. In particular, for the hypergraph bicoloring and coloring problems, its asymptotic expansion at large $k,q$ shows that the relevant scale of constraint densities (for the graph coloring problem we use the more natural average degree $c=2\alpha$) is
\beq
\alpha = \alpha(k,\g) = \frac{2^{k-1}}{k} (\ln k + \ln \ln k + \gamma) \ , \qquad
c= c(q,\g) = q (\ln q + \ln \ln q + \gamma) \ ,
\label{eq_scale}
\eeq
with $\gamma$ a finite constant parameter, the rigidity transition $\ar$ occuring on this scale when $\g$ crosses the critical value $\gr=1$ (for both models). It turns out that the asymptotic behavior of $\ad$ is also on the scale of Eq.~(\ref{eq_scale}), with another constant $\gd \le \gr$. This statement follows from a series of rigorous works on this problem: for the coloring problem \cite{Sly08} proved that $1-\ln 2 \le \gd \le \gr =1$, and the strict inequality $\gd < 1$ was later obtained in~\cite{SlyZhang16}, implying the asymptotic existence of a regime $[\gd, \gr]$ where reconstruction is possible but naive reconstruction is not. A large family of models, including the two discussed here, was also adressed in~\cite{MoReTe11_recclus} which proved  $\ad(k) \ge \frac{2^{k-1}}{k} \ln k$, $\cd(q) \ge q \ln q$, confirming the leading term in the scaling ~(\ref{eq_scale}). 

We report in this paper a study of the asymptotic expansion of $\ad$ for the graph coloring and hypergraph bicoloring problems. We will show that the transition indeed happens on the scale of constraint densities defined in (\ref{eq_scale}), with the same value of $\gd$ for both models, which is to some extent surprising given their different microscopic nature. We characterize $\gd$ in terms of the behavior of a functional equation, whose numerical study yields the estimate $\gd \approx 0.871$, while an analytical treatment provides the lowerbound $\gamma_{\rm d} \ge 1 + \ln (2 (\sqrt{2}-1)) \approx 0.812$. Even with the less demanding level of rigor of theoretical physics, to which we stick here, the computation is relatively involved because of the asymptotic proximity of the thresholds $\ad$ and $\ar$, at the origin of a quite singular behavior of the probability distributions describing the intermediate regime of reconstruction without naive reconstruction. This might be an explanation for some previous incorrect statements in the literature, in particular~\cite{MontanariRicci08} wrongly claimed that $\gd=\gr$ for the $k$-SAT problem. 

The rest of the paper is organized as follows. We start in Section~\ref{sec_definitions} by defining more explicitly the tree reconstruction problem and the basic equations that describe it. Then in Sec.~\ref{sec_largek} we perform the large $k$ limit on the hypergraph bicoloring model and reduce the determination of $\gd$ to the study of a reduced set of equations, in which $k$ does not appear anymore. This reduced problem takes the form of recursive equations on a sequence of probability distributions, determining a correlation function at distance $n$. The study of the large $n$ limit of this reduced problem, from which $\gd$ can be finally deduced, is performed in Sec.~\ref{sec_largen}; this limit requires some additional reparametrizations in the interesting regime. Finally we draw our conclusions and propose some perspectives for future works in Sec.~\ref{sec_conclu}. For simplicity in the main part of the text we concentrate on the hypergraph bicoloring case, and devote the Appendix~\ref{sec_coloring} to the graph coloring problem: we show that the large $q$ limit yields exactly the same reduced problem than the large $k$ limit of the hypergraph bicoloring, hence the study of Sec.~\ref{sec_largen} and the determination of $\gd$ is common to both.

\section{The tree reconstruction problem for finite $k$}
\label{sec_definitions}

\subsection{The recursive distributional equations for the tree reconstruction problem}

\begin{figure}
\begin{center}
\includegraphics[width=8cm]{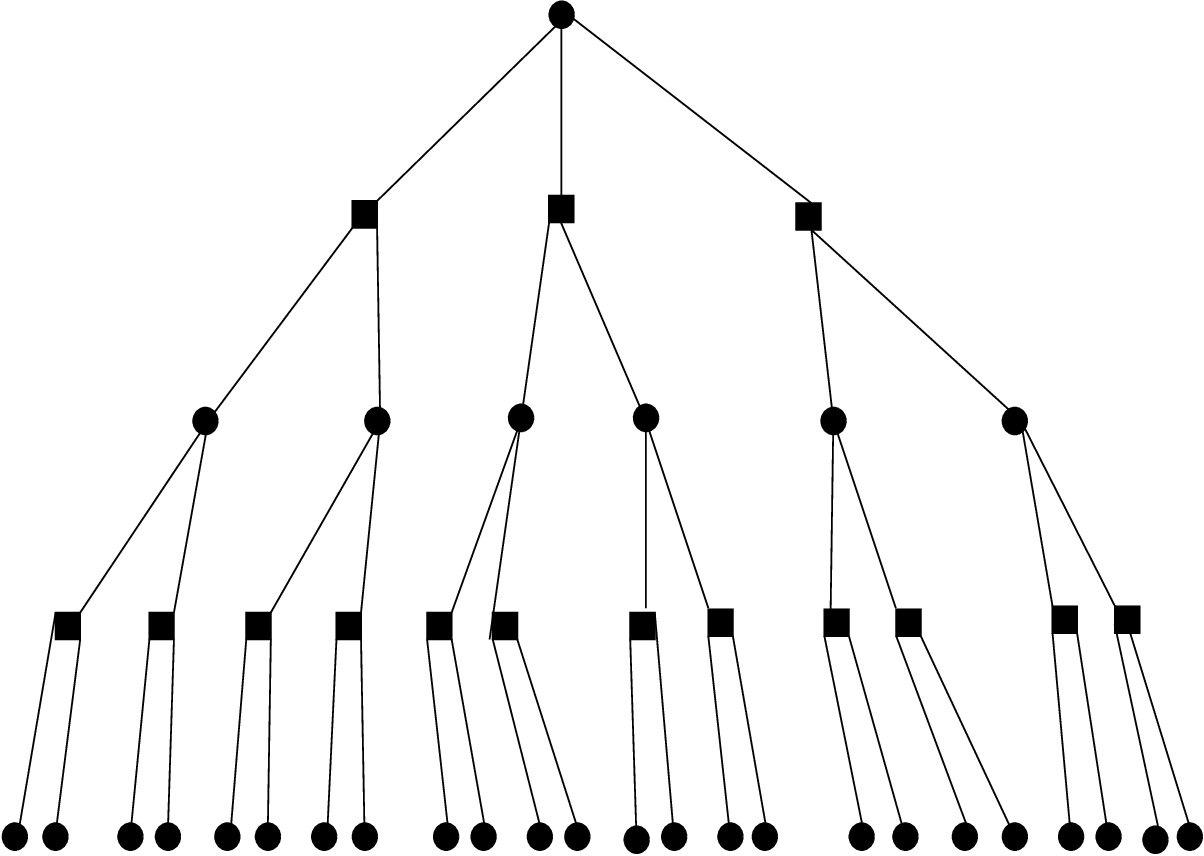}
\caption{An example of an hypertree, vertices being represented by circles, while hyperedges are drawn as squares linked to $k$ vertices, with $k=3$ on the figure.}
\label{fig_tree}
\end{center}
\end{figure}

As mentioned in the introduction the dynamic transition $\ad$ can be characterized in several ways; we shall exploit here the perspective provided by the tree reconstruction problem, that we now briefly describe in the specific case of the hypergraph bicoloring, referring the reader to~\cite{MoPe03,MezardMontanari06} for more generic and extensive discussions. Consider an hypertree $G=(V,E)$, with vertices $i \in V$, and hyperedges $a \in E$, where an hyperedge is a set denoted $\da$ of $k \ge 3$ vertices (see Fig.~\ref{fig_tree} for an example); we shall denote similarly $\di$ the set of hyperedges adjacent to a vertex $i$. An Ising spin variable $\s_i \in \{-1,+1\}$ is placed on each vertex $i \in V$ of the hypertree, with their global configuration denoted $\us = \{\s_i , i \in V \}$, the notation $\us_S= \{ \s_i , i \in S \}$ corresponding to the configuration on a subset $S \subset V$ of the vertices. A proper bicoloring of $G$ is defined as a configuration $\us$ such that no hyperedge is monochromatic, in other words every hyperedge must have among its $k$ neighboring vertices at least one spin equal to $+1$ and one equal to $-1$. The uniform distribution over these proper bicolorings, $\eta(\us)$, can thus be written as
\beq
\eta(\us) = \frac{1}{Z} \prod_{a \in E} w(\us_{\da}) \ , \qquad
w(\s_1,\dots,\s_k) = \ind(\s_1,\dots,\s_k \, \text{n.a.e.} ) = 1-\ind(\s_1=\dots=\s_k) \ , 
\eeq
where $Z$ is a normalizing factor counting the number of proper bicolorings, $\ind(A)$ is the indicator function of the event $A$, and n.a.e. is the abbreviation of not all equal.

Thanks to the invariance of $\eta$ under a global spin flip, $\eta(-\us) = \eta(\us)$, and to the acyclic property of the tree $G$, it is actually easy to sample configurations from $\eta$ by a recursive, broadcasting procedure described as follows. Choose an arbitrary vertex $i_0$ to be called the root of the tree (the one drawn at the top of Fig.~\ref{fig_tree}), and set its spin $\s_{i_0}$ to $\pm1$ with probability $1/2$. Then, independently for each edge $a$ adjacent to the root, draw the configuration of the $k-1$ other variables with the conditional probability $p(\us_{\dami_0}|\s_{i_0})$, where
\beq
p(\s_1,\dots,\s_{k-1}|\s) = \frac{1}{2^{k-1}-1} (1-\ind(\s_1=\dots=\s_{k-1}=\s)) \ .
\label{eq_p_broadcast}
\eeq
Once the values of all the spins at distance 1 from the root have been set in this way the same process can be iterated, each of these vertices of the first generation being in turn considered as the root of the subtree lying below it.

This description of the sampling from $\eta$ as a broadcast process naturally calls for an interpretation as the transmission of an information, the value of the spin at the root, through noisy channels, the hyperedges, towards a set of receivers, the vertices at a certain distance $n$ from the root, to be denoted $B_n$. In this information theoretic perspective the tree reconstruction problem asks the following question: given only the values $\us_{B_n}$ of the spins at distance $n$ from the root in a configuration $\us$ generated as above, how much information is available on the value of the root $\s_{i_0}$, and does a non-vanishing amount of information on $\s_{i_0}$ survives in the $n \to \infty$ limit ? This question is equivalent to the existence of  long-range non-trivial point-to-set correlations under the probability measure $\eta$, the ``point'' being the root and the ``set'' the vertices at distance $n$ from it. In this Bayesian setting the information-theoretical optimal strategy of an observer having to reconstruct $\s_{i_0}$ from $\us_{B_n}$ is to compute the posterior probability $\eta(\s_{i_0}|\us_{B_n})$. Thanks to the tree structure of the interaction graph this strategy is actually computationally feasible and can be performed in a recursive way. Consider indeed the so-called Belief Propagation (BP) equations, also known as message passing or dynamic programming algorithm~\cite{KschischangFrey01}, bearing upon probability measures over $\{-1,+1\}$, $\eta_{i \to a}$, $\heta_{a \to i}$ for each edge between an hyperedge $a$ and an adjacent vertex $i \in \da$:
\beq
\eta_{i \to a}(\tau_i) = \frac{1}{z_{i \to a}} \prod_{b \in \dima} \heta_{b \to i}(\tau_i) \ , \quad
\heta_{a \to i}(\tau_i) = \frac{1}{z_{a \to i}} \sum_{\ut_{\dami}} w(\tau_i,\ut_{\dami}) \prod_{j \in \dami} \eta_{j \to a}(\tau_j) \ ,
\label{eq_BP_eta}
\eeq
where $z_{i \to a}$ and $z_{a \to i}$ are chosen to enforce the normalization of these probability laws. A moment of thought reveals that, once supplemented by the boundary conditions
\beq
\eta_{i \to a}(\tau_i) = \delta_{\tau_i,\s_i} \ \ \text{for} \ \ i \in B_n \ ,
\label{eq_boundary_eta}
\eeq
these BP equations admit a unique solution for any finite tree, and that the posterior probabilities $\eta(\s_i|\us_{B_n})$ can be determined, for $i \notin B_n$, from these messages according to
\beq
\eta(\s_i|\us_{B_n}) = \frac{1}{z_i} \prod_{a \in \di} \heta_{a \to i}(\s_i) \ ,
\eeq
with again $z_i$ a normalizing constant; note that in order to lighten the notations we have kept implicit the dependency of the messages on the observed variables $\us_{B_n}$. As the messages $\eta_{i \to a}$ and $\heta_{a \to i}$ are probability distributions over a binary variable they can be parametrized by a single real number, that we shall denote $h_{i \to a}$ and $u_{a \to i}$ respectively, and define by
\beq
\eta_{i \to a}(\tau_i) = \frac{1+h_{i \to a} \tau_i}{2} \ , \qquad
\heta_{a \to i}(\tau_i) = \frac{1+u_{a \to i}\tau_i}{2} \ .
\eeq
These numbers $h_{i \to a}$ and $u_{a \to i}$ are the average values of $\tau_i$ under the laws $\eta_{i \to a}$ and $\heta_{a \to i}$, they are thus in $[-1,1]$ and correspond to the magnetization of the spin; with a slight abuse of vocabulary we shall sometimes call them ``fields'' in the following. With this parametrization, and for the special case of the hypergraph bicoloring problem, the BP equations (\ref{eq_BP_eta}) become
\beq
h_{i \to a}= f(\{u_{b \to i}\}_{b \in \dima}) \ , \qquad
u_{a \to i}= g(\{h_{j \to a}\}_{j \in \dami}) \ , 
\eeq
with the functions $f$ and $g$ defined by
\beq
f(u_1,\dots,u_l) = \frac{\underset{i=1}{\overset{l}{\prod}} \frac{1+u_i}{2} - \underset{i=1}{\overset{l}{\prod}} \frac{1-u_i}{2} }{\underset{i=1}{\overset{l}{\prod}}  \frac{1+u_i}{2} + \underset{i=1}{\overset{l}{\prod}} \frac{1-u_i}{2}} \ ,
\qquad
g(h_1,\dots,h_{k-1}) = \frac{\underset{i=1}{\overset{k-1}{\prod}} \frac{1-h_i}{2} - \underset{i=1}{\overset{k-1}{\prod}} \frac{1+h_i}{2} }{2 -  \underset{i=1}{\overset{k-1}{\prod}} \frac{1+h_i}{2} -  \underset{i=1}{\overset{k-1}{\prod}} \frac{1-h_i}{2}} \ .
\label{eq_defs_fg}
\eeq
The boundary condition (\ref{eq_boundary_eta}) becomes
\beq
h_{i \to a} = \s_i \ \ \text{for} \ \ i \in B_n \ ,
\eeq
and the average value of the spin at the root, in the posterior probability measure conditional on the observations of the spins at distance $n$, reads
\beq
h_{i_0}=\sum_{\s_{i_0}} \s_{i_0}  \eta(\s_{i_0}|\us_{B_n}) = f(\{u_{a \to i_0}\}_{a \in \di_0}) \ .
\eeq
It is this value which has to be compared with the actual value of $\s_{i_0}$ in the broadcast process that led to $\us_{B_n}$ in order to assess the amount of information transmitted between $i_0$ and $B_n$; once averaged over $\eta(\us)$ this yields the point-to-set correlation function.

The computation descrived above was performed for a given tree. Suppose now that the tree itself is random, and drawn as an instance of a Galton-Watson process with $n$ generations, in which every vertex has a random offspring (number of descendent hyperedges) distributed as a Poisson law of parameter denoted $\alpha k$. Let us call as above $h \in [-1,1]$ the posterior magnetization of the root given the observations of the spins on the vertices of the $n$-th generation (the root corresponding to the generation $n=0$). $h$ is a random variable because of the randomness in the generation of the Galton-Watson tree on the one hand, and in the broadcast process yielding $\us_{B_n}$ on the other hand. We shall denote $P_{\s,n}(h)$ its distribution when the broadcast is conditioned to the root being equal to $\s$. As both the broadcast process (from the root downwards) and the resolution of the BP equations (from the leaves upwards) decompose recursively along the branches of the tree, it is possible to obtain from the above analysis an inductive formula relating these distributions for trees of depth $n$ and $n+1$, namely
\begin{align}
  P_{\s,n+1}(h) &= \sum_{l=0}^\infty e^{-\alpha k} \frac{(\alpha k)^l}{l!} \int \prod_{i=1}^l \dd \hP_{\s,n}(u_i) \, \delta(h-f(u_1,\dots,u_l)) \ , \label{1RSBeq1_sigma} \\
\hP_{\s,n}(u) &= \sum_{\s_1,\dots,\s_{k-1}} p(\s_1,\dots,\s_{k-1}|\s) \int \prod_{i=1}^{k-1} \dd P_{\s_i,n}(h_i) \, \delta(u-g(h_1,\dots,h_{k-1})) \ ,
\label{1RSBeq2_sigma}
\end{align}
with the initial condition $P_{\s,0}(h) = \delta(h-\s)$ that expresses the observation of the variables at the boundary of the tree. These are examples of Recursive Distributional Equations (RDEs), as they define by recursion a sequence of probability distributions. They can be equivalently written as equalities in distribution between random variables, for instance (\ref{1RSBeq1_sigma}) means $h \eqd f(u_1,\dots,u_l)$ with $h$ drawn from $P_{\s,n+1}$, $l$ is a Poisson distributed random variable with mean $\alpha k$, and the $u_i$'s are i.i.d. copies of a random variable of law $\hP_{\s,n}$.

One can slightly simplify these equations by noticing that the invariance of the bicoloring problem under a global spin-flip of all its variables implies that $P_{-\s,n}(h)=P_{\s,n}(-h)$ and $\hP_{-\s,n}(u)=\hP_{\s,n}(-u)$; this can be checked by induction from (\ref{1RSBeq1_sigma},\ref{1RSBeq2_sigma}), using $p(\s_1,\dots,\s_{k-1}|\s)=p(-\s_1,\dots,-\s_{k-1}|-\s)$ and the fact that $f$ and $g$ change sign when all their arguments are multiplied by $-1$. It is thus redundant to track the evolution with $n$ of the distributions with both $\s=+1$ and $\s=-1$, we shall instead define $P_n(h)=P_{+1,n}(h)$ and $\hP_n(u)=\hP_{+1,n}(u)$ and close the equations on these two sequences of distributions:
\begin{align}
\label{1RSBeq1}
  P_{n+1}(h) &= \sum_{l=0}^\infty e^{-\alpha k} \frac{(\alpha k)^l}{l!} \int \prod_{i=1}^l \dd \hP_n(u_i) \, \delta(h-f(u_1,\dots,u_l)) \ , \\
\hP_n(u) &= \sum_{\s_1,\dots,\s_{k-1}} p(\s_1,\dots,\s_{k-1}|+1) \int \prod_{i=1}^{k-1} \dd P_n(h_i) \, \delta(u-g(h_1\s_1,\dots,h_{k-1}\s_{k-1})) \ .
\label{1RSBeq2}
\end{align}

The answer to the reconstructibility question raised above can be read off from the behavior of $P_n(h)$ in the large $n$ limit: if it tends to the trivial distribution $\delta(h)$ (which is always a stationary solution of (\ref{1RSBeq1},\ref{1RSBeq2})), then all information on the value of the root has been washed out and the reconstruction problem is not solvable. On the contrary if the limit of $P_n(h)$ is non-trivial then the observation of $\us_{B_n}$ contains some information on the root and the problem is said to be solvable. The occurence of these two situations depend on the parameters $k$ and $\alpha$ of the model; increasing $\alpha$ gives rise to a larger number of spin variables to be observed, which makes the inference problem easier. Hence it is natural to expect the existence of a threshold $\ad(k)$ such that the problem is unsolvable (resp. solvable) for $\alpha < \ad(k)$ (resp. $\alpha > \ad(k)$).

It is actually more convenient to describe this transition by a scalar order parameter (instead of the functional one $P_n$), the point-to-set correlation function that we shall define as
\beq
C_n(\alpha,k)=\int \dd P_n(h) \, h \ , \qquad C(\alpha,k)=\lim_{n\to \infty} C_n(\alpha,k) \ .
\eeq
This is indeed an order parameter for the transition in the sense that $C(\alpha,k)>0$ if and only if $\alpha > \ad(k)$. This equivalence is a consequence of some symmetry properties of the distributions $P_n$ that we will describe now.

Let us call $T_n(h)$ the distribution of the posterior magnetization $h$ of the root, in a broadcast process which is not conditioned on the value of the root in $\us$, i.e. $T_n(h)=(P_{+,n}(h)+P_{-,n}(h))/2$. A consequence of Bayes theorem applied to the joint law between the spins at the root and on the boundary vertices is the following converse relation between conditional and unconditional laws, $P_{\s,n}(h)=(1+\s h)T_n(h)$ (see~\cite{MezardMontanari06,MoReTe11_recclus,RichardsonUrbanke07} for more details on this property and its consequences). In addition the invariance of the problem under a global spin-flip implies that $T_n(-h)=T_n(h)$, which yields a symmetry constraint on the distributions $P_n$,
\beq
P_n(-h)=\frac{1-h}{1+h} \, P_n(h) \ , \qquad \int \dd P_n(h) \, b(h) = \int \dd P_n(h) \, b(-h) \, \frac{1-h}{1+h} \ ,
\label{eq_Bayes}
\eeq
the second equality being valid for any function $b(h)$ such that the integrals exist. This symmetry (sometimes called Nishimori symmetry~\cite{NishimoriBook01}, and also fulfilled by $\hP_n(u)$) implies several identities between the moments of $h$ that can be derived by appropriate choices of the test function $b$~\cite{MoReTe11_recclus,RichardsonUrbanke07}; here we shall only state the simplest one, namely
\beq
C_n(\alpha,k)=\int \dd P_n(h) \, h = \int \dd P_n(h) \, h^2 \ , 
\label{eq_Bayes_moment}
\eeq
that can be obtained from (\ref{eq_Bayes}) with $b(h)=h(1-h)$. This is enough to justify our statement of the characterization of the reconstruction transition via the behavior of $C_n$: if the latter vanishes in the $n \to \infty$ limit this implies that both the average and the variance of $P_n$ go to zero, hence $P_n$ tends to the trivial distribution $\delta(h)$.

 \subsection{Numerical resolution for finite $k$}
\label{sec_num_finitek}
The distributional equations (\ref{1RSBeq1},\ref{1RSBeq2}) can be solved numerically relatively easily for finite values of $k$ by using a population dynamics algorithm~\cite{ACTA73,MezardParisi01}, whose main idea is to approximate a probability distribution by the empirical distribution of a large sample of representative elements. Suppose indeed that at some step $n$ one has an approximation of $P_n(h)$ written as
\begin{align}
\label{population}
P_n(h) \approx \frac{1}{{\cal N}} \sum_{i=1}^{\cal N}\delta(h-h_i^{(n)}) \ ,
\end{align}
where ${\cal N} \gg 1$ is the size of the population, that controls the numerical accuracy of the procedure. One can then insert this form in the r.h.s. of (\ref{1RSBeq2}) to obtain an approximation of $\hP_n(u)$ as
\begin{align}
\hP_n(u) \approx \frac{1}{{\cal N}} \sum_{i=1}^{\cal N}\delta(u-u_i^{(n)}) \ ,
\end{align}
where each of the representants $u_i^{(n)}$ has been constructed independently by drawing $\s_1,\dots,\s_{k-1}$ according to the law $p(\s_1,\dots,\s_{k-1}|+)$, then $k-1$ indices $i_1,\dots,i_{k-1}$ uniformly at random in $\{1,\dots,\cal N\}$, and setting $u_i^{(n)}=g\left(\s_1 h_{i_1}^{(n)},\dots,\s_{k-1} h_{i_{k-1}}^{(n)}\right)$. Similarly, $P_{n+1}(h)$ can be approximated by an empirical distribution of the form (\ref{population}), where according to (\ref{1RSBeq1})  each of the representants $h_i^{(n+1)}$ is obtained by drawing an integer $l$ from the Poisson law $p_l=e^{-\alpha k}\frac{(\alpha k)^l}{l!}$, then $l$ indices $i_1,\dots,i_l$ uniformly at random in $\{1,\dots,\cal N\}$, and taking $h_i^{(n+1)}=f\left(u_{i_1}^{(n)},\dots,u_{i_l}^{(n)}\right)$. The initial condition $P_0(h)=\delta(h-1)$ can obviously be represented by a sample with all representants $h_i^{(0)}$ equal to 1, and at every iteration step observables can be estimated as empirical averages,
\beq
\int \dd P_n(h) F(h) \approx \frac{1}{{\cal N}} \sum_{i=1}^{\cal N} F(h_i^{(n)})
\eeq
for an arbitrary function $F$.

\begin{figure}
\begin{center}
  \includegraphics[width=8cm]{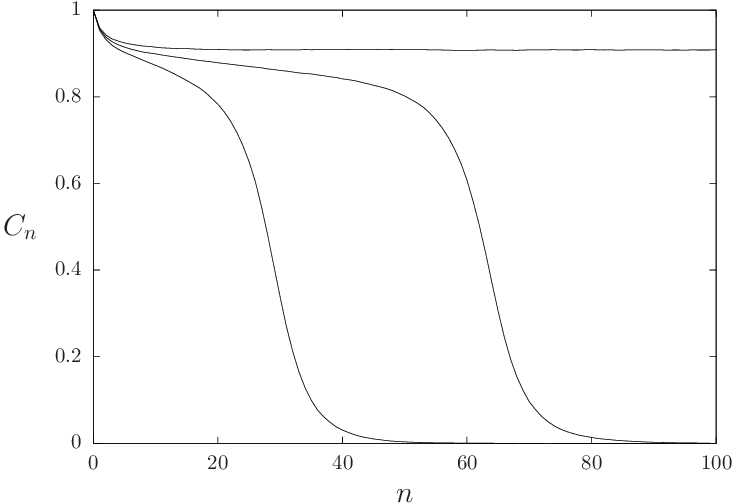}
  \hspace{1cm}
  \includegraphics[width=8cm]{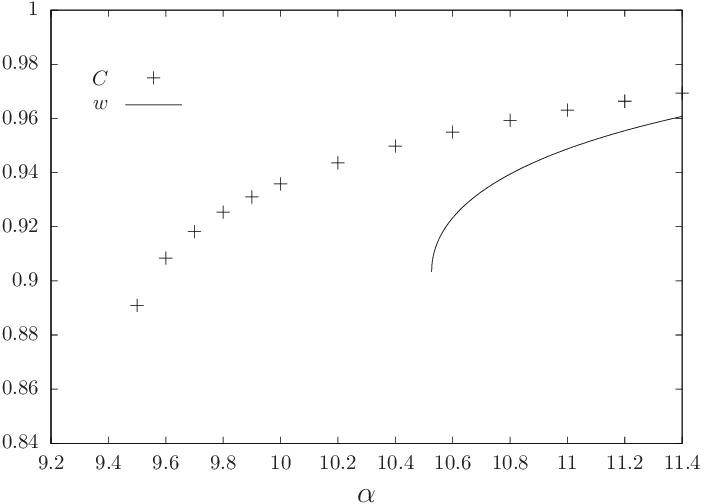}
  \caption{Left: the correlation function $C_n(\alpha,k)$ as a function of $n$ for $k=5$ and from left to right $\alpha=9.2$, $\alpha=9.4$, $\alpha=9.6$.
    Right: the large $n$ limit $C(\alpha,k)=\lim_n C_n(\alpha,k)$ as a function of $\alpha$ for $k=5$, along with the contribution of the hard fields $w(\alpha,k)=\lim_n w_n(\alpha,k)$ discussed in Sec.~\ref{sec_naive}. The first curve is non zero for $\alpha \ge \ad(k=5) \approx 9.465$, the second one for $\alpha \ge \ar(k=5) \approx 10.526$.  }
\label{fig_k5}
\end{center}
\end{figure}

The figure~\ref{fig_k5} presents some numerical results obtained with this procedure. The left panel displays the decay of the correlation function $C_n(\alpha,k)$ as a function of the distance $n$ for different values of $\alpha$ and $k=5$ (all values of $k$ large enough are qualitatively similar, but small values of $k$ behaves differently, see for instance~\cite{gabrie2017phase}). When $\alpha$ is small this function decays to 0 as $n \to \infty$, signalling the impossibility of reconstruction; when $\alpha$ increases the decay gets slower and proceeds in two steps, with a longer and longer plateau at a strictly positive value developping as $\alpha$ gets closer to the transition; finally for $\alpha > \ad$ the plateau lasts forever, the large $n$ limit $C(\alpha,k)$ is strictly positive. This quantity, that jumps discontinuously from 0 to a strictly positive value when $\alpha$ crosses $\ad$, is plotted as a function of $\alpha$ in the right panel of Fig.~\ref{fig_k5}. To increase the numerical accuracy we estimated the limit by averaging the value of $C_n$ when $n$ is large enough to have reached its plateau behavior.

\subsection{The naive reconstruction procedure}
\label{sec_naive}

The definition of the reconstruction transition $\ad$ and its analysis in terms of the RDE on $P_n$ described above are based on the optimal inference algorithm, namely the computation of the posterior probability $\eta(\s_i|\us_{B_n})$ via the BP equations (\ref{eq_BP_eta}). One can nevertheless consider a simpler, suboptimal inference procedure, that aims at answering the following question: is $\s_i$ uniquely determined in the measure $\eta(\cdot|\us_{B_n})$, in other words is $\s_i$ constant in all the proper bicolorings of the graphs that take the values $\us_{B_n}$ on the boundary vertices? It is clear that if the answer is yes then the inferred value of $\s_i$ is the correct one that was used in the broadcast, hence if reconstruction is possible in this strong sense of certain inference it is also possible in the definition introduced above.

We shall now determine the probability of success of this naive reconstruction procedure, that corresponds to a projection of the BP algorithm towards its Warning Propagation~\cite{BraunsteinMezard05} version that only keeps sure beliefs and discards partially biased ones. To incorporate naturally this computation in the one presented above we decompose the field distributions as follows:
\beq
P_n(h) = w_n \, \delta(h-1) + (1-w_n) \, Q_n(h) \ , \qquad
\hP_n(u) = \hw_n \, \delta(u-1) + (1-\hw_n) \, \hQ_n(u) \ ,
\label{eq_bicol_def_hard}
\eeq
where $Q_n$ and $\hQ_n$ are probability measures with no atom in $1$. The weight $w_n$ of the ``hard field'' $h=1$, that constrain completely the variable at the root and makes it a frozen variable under the boundary condition $\us_{B_n}$, is precisely the probability of success of the naive reconstruction procedure. In the following we shall call $Q_n$ and $\hQ_n$ the distributions of soft fields; note that they do not contain atoms in $-1$ (because a variable $\s_i$ cannot be forced by $\us_{B_n}$ to another value that it had in the broadcast), and that they enjoy the same Bayes symmetry (\ref{eq_Bayes}) as the complete distributions $P_n$ and $\hP_n$. Inserting this decomposition in (\ref{1RSBeq1},\ref{1RSBeq2}), and considering the possible combinations of arguments of the functions $f,g$ in (\ref{eq_defs_fg}) that yields $h,u=1$ one easily obtains the following evolution equations for $w_n$ and $\hw_n$:
\beq
w_{n+1} = 1-e^{-\alpha k \hw_n} \ , \qquad \hw_n = \frac{w_n^{k-1}}{2^{k-1}-1} \ .
\label{eq_wn_hwn}
\eeq
The first one expresses the fact that a spin is perfectly recovered as soon as one of its neighboring interactions forces it to its correct value, while the second one shows that this latter event happens when in the broadcast the $k-1$ variables adjacent to it have been given the same value, and that they all have been perfectly recovered.

Assembling these two equations we obtain a simple recursive equation on $w_n$,
\beq
w_{n+1} = 1-e^{- \Gamma(\alpha,k) w_n^{k-1} } \ ,
\qquad \Gamma(\alpha,k) = \frac{\alpha k}{2^{k-1}-1} \ ,
\label{eq_recursion_wn}
\eeq
with the initial condition $w_0=1$. By a numerical inspection of the shape of this recursion function one easily realizes that for $k \ge 3$ the fixed point reached by $w_n$ for $n \to \infty$ undergoes a discontinuous bifurcation from zero to a strictly positive value when $\Gamma$ crosses a critical value $\Gamma_{\rm r} = \Gamma_{\rm r}(k)$. The latter can be determined by noting that at such a bifurcation the derivative of the recursion function must be equal to 1, hence that $\Gamma_{\rm r}$ and $w_{\rm r}$, the fixed point at the bifurcation, are solutions of
\beq
w_{\rm r} = 1-e^{- \Gamma_{\rm r} w_{\rm r}^{k-1}} \ , \qquad 1 = (k-1) w_{\rm r}^{k-2} \Gamma_{\rm r} e^{- \Gamma_{\rm r} w_{\rm r}^{k-1}} \ .
\label{eq_wr_Gammar}
\eeq
One can close these two equations on a single one that determines $w_{\rm r}$: $1=(k-1)\ln(1-w_{\rm r})(1-1/w_{\rm r})$, from which one obtain $\Gamma_{\rm r}(k)=-\ln(1-w_{\rm r})/(w_{\rm r}^{k-1})$, and in terms of the original parameter $\alpha$ this ``rigidity'' transition occurs at $\ar(k) = (2^{k-1}-1) \Gamma_{\rm r}(k) / k$. 

\begin{figure}
\begin{center}
\includegraphics[width=8cm]{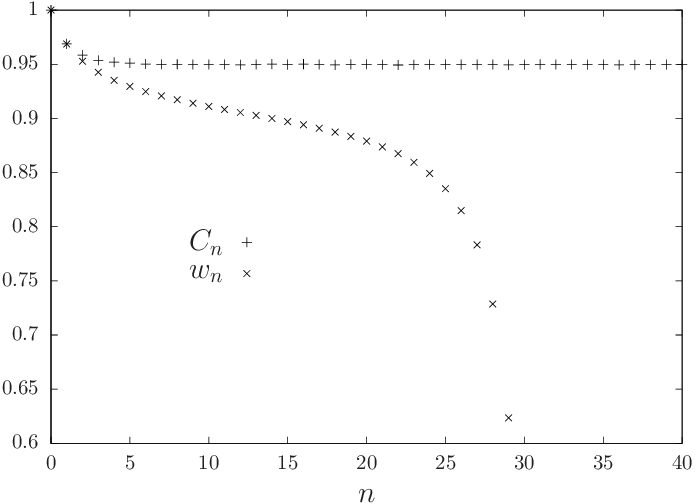}
\caption{The evolution of $C_n$ and $w_n$ for $k=5$, $\alpha =10.4$, in the intermediate regime between $\ad$ and $\ar$: the fraction $w_n$ of hard fields falls to zero (the vertical range has been reduced for the sake of readability) while the correlation function $C_n$ tends to a positive constant.}
\label{fig_k5_a10.40}
\end{center}
\end{figure}

From the intuitive interpretation of this computation as the analysis of a suboptimal reconstruction algorithm it is clear that $\ar$ should be an upperbound on $\ad$, and indeed if $w=\lim_n w_n$ is strictly positive then $P_n$ cannot converge to $\delta(h)$. Moreover this probability $w$ of perfect reconstruction is a lowerbound on the correlation function $C$, as for any value of $n$ the fraction of hard fields $w_n$ is a lowerbound on $C_n$:
\beq
C_n(\alpha,k)=\int \dd P_n(h) h = w_n + (1-w_n) \int \dd Q_n(h) h \ , \qquad
\text{and} \ \ \int \dd Q_n(h) h = \int \dd Q_n(h) h^2 \ge 0 \ ,
\label{eq_Cn_decomposed}
\eeq
because of the consequence stated in (\ref{eq_Bayes_moment}) of the Bayes symmetry, that is also enjoyed by $Q_n$. An illustration of the bound $w(\alpha,k) \le C(\alpha,k)$ can be found for $k=5$ in the right panel of Fig.~\ref{fig_k5}. For finite $k$ this bound, as well as its consequence $\ad(k) \le \ar(k)$, are not tight. There is an intermediate regime $\ad < \alpha < \ar$ where reconstruction is possible but naive reconstruction is not, all the relevant information on the value of the root is asymptotically contained in the soft fields distribution. Another view on this phenomenon is given in Fig.~\ref{fig_k5_a10.40} where one sees the fraction $w_n$ of hard fields fall to zero as $n$ grows while $C_n$ remains at a positive plateau value. Also when $\alpha > \ar$ one can see on the right panel of Fig.~\ref{fig_k5} that $C > w$ for $k=5$, i.e. the soft fields do bring some additional information on the value of the root even in the presence of hard fields.

\subsection{The distribution of the soft fields}

We will now derive the recursion equations on the distribution of the soft fields $Q_n$ and $\hQ_n$ introduced in (\ref{eq_bicol_def_hard}), that complete the equations (\ref{eq_wn_hwn}) for the evolution of the weights $w_n$ and $\hw_n$ of hard fields. Both are obtained by plugging the decomposition (\ref{eq_bicol_def_hard}) into the recursion equations (\ref{1RSBeq1},\ref{1RSBeq2}) on $P_n$ and $\hP_n$.

As $f(u_1,\dots,u_l)=1$ as soon as one of the arguments is equal to 1, i.e. as soon as one of the neighboring interactions forces the root variable, it is easy to see from (\ref{1RSBeq1}) that the soft part of $P_n$ arises from the combination of only soft $u$'s, namely 
\beq
Q_{n+1}(h) = \sum_{l=0}^\infty e^{-\alpha k (1-\hw_n)} \frac{(\alpha k(1-\hw_n))^l}{l!} \int \prod_{i=1}^l \dd \hQ_n(u_i) \, \delta(h-f(u_1,\dots,u_l)) \ .
\label{eq_recursion_Qn}
\eeq

The treatment of (\ref{1RSBeq2}) will require a little bit more work in order to put the resulting equation into a form convenient for the rest of the paper. Denoting $p$ the number of soft fields picked in the r.h.s. of (\ref{1RSBeq2}), we first rewrite this equation as
\begin{align}
\hP_n(u) = \sum_{p=0}^{k-1} \binom{k-1}{p} w_n^{k-1-p}(1-w_n)^p & \sum_{\s_1\dots \s_{k-1}}p(\s_1\dots \s_{k-1}|+) \nonumber \\ & \int \dd Q_n(h_1)\dots \dd Q_n(h_p) \, \delta(u-g(\s_1 h_1\dots \s_p h_p,\s_{p+1}\dots ,\s_{k-1})) \ .
\end{align}
As explained above the term $\delta(u-1)$ arises solely from the term $p=0$, $(\s_1\dots \s_{k-1})=(-,\dots,-)$. Moreover one realizes by inspection of the expression of $g$ in (\ref{eq_defs_fg}) that $u=0$ as soon as among the hard fields $(\s_{p+1}\dots ,\s_{k-1})$ at least one is positive and at least one is negative; this expresses the fact that in such a situation the bicoloring condition is satisfied whatever the value of the spin at the root. Combining these observations with the expression (\ref{eq_p_broadcast}) of the broadcasting probability $p(\s_1\dots \s_{k-1}|+)$ one obtains after a short computation the following expression for the soft part of $\hP_n$,
\begin{align}
\nonumber 
(1-\hw_n)\hQ_n(u) &= \frac{2^{k-1}}{2^{k-1} -1} \left[1 + (1-w_n)^{k-1} - 2 \left(1-\frac{w_n}{2} \right)^{k-1} \right] \delta(u) \\
\nonumber 
&+ \sum_{p=1}^{k-2} \binom{k-1}{p} (1-w_n)^p w_n^{k-1-p}
\int \prod_{i=1}^p \dd Q_n(h_i) \frac{1}{2^{k-1}-1}  \sum_{\s_1,\dots,\s_p} \delta(u-g_p(\s_1 h_1,\dots,\s_p h_p))  \\
\nonumber 
&+  \sum_{p=1}^{k-2} \binom{k-1}{p} (1-w_n)^p w_n^{k-1-p}
\int \prod_{i=1}^p \dd Q_n(h_i) \frac{1}{2^{k-1}-1} \underset {\s_1,\dots,\s_p}{{\sum}'} \delta(u+g_p(\s_1 h_1,\dots,\s_p h_p)) \\
& +\frac{(1-w_n)^{k-1}}{2^{k-1}-1} \underset {\s_1,\dots,\s_{k-1}}{{\sum}'} \int \prod_{i=1}^{k-1} \dd Q_n(h_i) \delta(u-g(\s_1 h_1,\dots,\s_{k-1} h_{k-1})) \ ,
\label{eq_recursion_hatQn}
\end{align}
 where the primed sums $\sum'$ exclude the configuration with all the spin arguments equal to $+1$, and where $g_p$ is defined as 
 \beq
g_p(h_1,\dots,h_p) =-g(h_1,\dots,h_p,+1,\dots,+1) = \frac{\underset{i=1}{\overset{p}{\prod}} \frac{1+h_i}{2}}{2 - \underset{i=1}{\overset{p}{\prod}} \frac{1+h_i}{2} } \ .
\label{eq_gp}
\eeq
The initial condition for this evolution of the soft fields distribution turns out to be $Q_1(h)=\delta(h)$, as can be easily realized by an explicit computation of $\hP_0(u)$ and $P_1(h)$ starting from $P_0(h)=\delta(h-1)$.

\section{The large $k$ limit for a finite distance $n$}
\label{sec_largek}

\subsection{Evolution of the hard fields}

\begin{figure}
\begin{center}
\includegraphics[width=8cm]{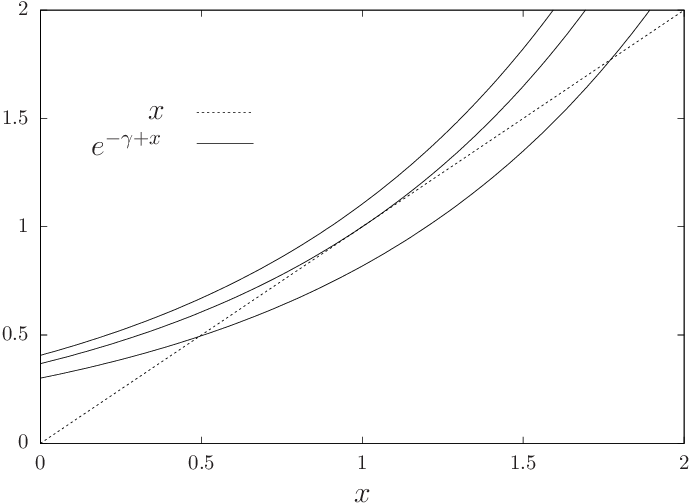}\hspace{1cm}
\includegraphics[width=8cm]{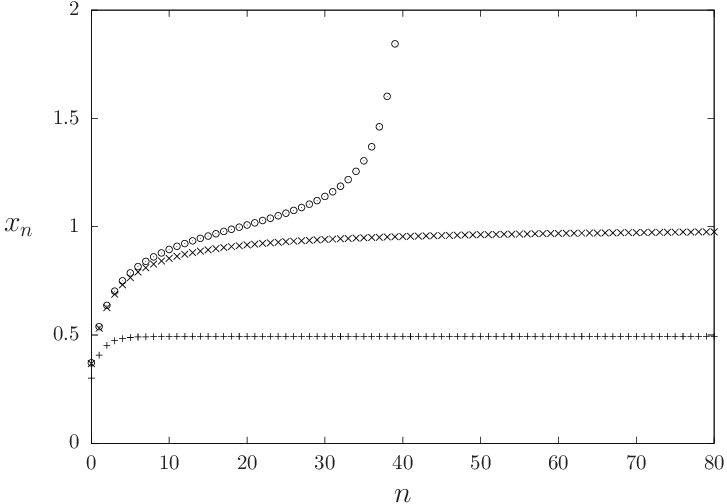}
\caption{Left: the recursion function $e^{-\g + x}$ (solid lines) for three values of $\g$ (from top to bottom $\g=0.9$, $\g=1=\gr$, $\g=1.2$) compared to $x$ (dashed line). Right: the iterates $x_n$ defined by $x_0=0$, $x_{n+1} = e^{- \g + x_n}$ for three values of $\g$ (from top to bottom $\g=0.99$, $\g=1=\gr$, $\g=1.2$).}
\label{fig_recurs_xn}
\end{center}
\end{figure}

We turn now to the first steps towards the main goal of this article, namely the determination of the asymptotic expansion of $\ad(k)$ when $k \to \infty$. Before confronting this functional bifurcation threshold it is much easier to consider the rigidity transition $\ar(k)$, that corresponds to the bifurcation of the scalar recursion equation (\ref{eq_recursion_wn}). We have given in (\ref{eq_wr_Gammar}) the expressions that determine analytically $\ar(k)$ for all $k$. The value of $\Gamma_{\rm r}(k)$ cannot be given explicitly when $k$ is finite, but it is easy to perform its asymptotic expansion at large $k$, which yields
\beq
\ar(k) = \frac{2^{k-1}}{k} (\ln k + \ln \ln k + \gr + o(1)) \ , \qquad
\text{with} \ \ \gr = 1 \ .
\label{eq_ar_expansion}
\eeq
As explained above this is an upperbound on the sought for dynamic transition $\ad(k)$, and the rigorous results of~\cite{MoReTe11_recclus} suggest (or prove for the coloring problem~\cite{Sly08,SlyZhang16}) that the asymptotic expansion of $\ad(k)$ has the same form as the one of $\ar(k)$, with a different constant term $\gd$.

We shall thus study the large $k$ limit taking $\alpha$ to diverge simultaneously with $k$ according to the function $\alpha(k,\gamma)$ defined in Eq.~(\ref{eq_scale}), with $\gamma$ a finite constant that becomes our control parameter for the density of constraints in this limit. It is instructive to investigate the behavior of the hard field weights $w_n$, solution of the recursion equation (\ref{eq_recursion_wn}), for a finite distance $n$. With this choice for the scaling of $\alpha$ with $k$ one easily obtains by induction on $n$ that the leading behavior of $w_n$ reads
\beq
w_n = 1 - \frac{x_n}{k \ln k} + o\left(\frac{1}{k \ln k} \right) \ ,
\label{eq_hard_largek}
\eeq
where $x_n$ is a $\g$-dependent sequence, defined recursively by
\beq
x_0 = 0 \ , \qquad x_{n+1} = e^{- \g + x_n} \ .
\label{eq_xn_first}
\eeq
This recursion function is illustrated on Fig.~\ref{fig_recurs_xn}. Its study is very simple, and unveils the following behavior for $x_n$: if $\gamma \ge 1$ then $x_n$ converges at large $n$ to a finite fixed point (depicted on the right panel of Fig.~\ref{fig_largek}), while for $\gamma < 1$ the sequence $x_n$ diverges as $n \to \infty$, very quickly (as iterated exponentials). The transition between these two behaviors thus occurs at the critical value of the parameter $\g = \gr =1$, which corresponds to the expansion of $\ar(k)$ stated above. Note that this coincidence is not as trivial as it may look at first sight: the expansion of Eq.~(\ref{eq_ar_expansion}) was obtained by first taking the limit $n\to\infty$, then $k \to \infty$, while in (\ref{eq_xn_first}) we have taken the limit $k \to \infty$, and only later studied the behavior of $x_n$ for large $n$. The commutativity of these two limits can in this case be traced back to the simple shape of the recursion equation (\ref{eq_recursion_wn}): either it has no non-trivial fixed point, or if it has one in the large $k$ limit it must be of order $1-O(1)/(k \ln k)$, and hence captured by the scaling (\ref{eq_hard_largek}).

\subsection{The reduced order parameter}
\label{sec_reduced}

Our determination of the asymptotic behavior of $\ad(k)$ will be based on three hypotheses that we now spell out explicitly. First of all, we assume that this transition occurs on the scale (\ref{eq_scale}), hence that the sought for constant $\gd$ can be defined as 
\beq
\gd = \lim_{k \to \infty} \left[ \frac{k \ad(k)}{2^{k-1}} - \ln k - \ln \ln k \right] \ ;
\eeq
as mentioned above the rigorous results of~\cite{Sly08,MoReTe11_recclus,SlyZhang16} support this hypothesis. We shall furthermore assume that on this scale of $\alpha$ the order parameter $C(\alpha,k)$ is either equal to 0 (for $\alpha < \ad(k)$) or scales with $k$ in the same way as its lowerbound provided by the hard field weight $w(\alpha,k)$ (see Eq.~(\ref{eq_hard_largek})), hence define a reduced order parameter as
\beq
\tC(\g) = \lim_{k \to \infty}  (1-C(\alpha(k,\g),k))k \ln k \ ,
\label{eq_def_tC_first}
\eeq
that should be finite for $\gamma > \gd$, and diverge to $+\infty$ for $\gamma < \gd$. Note that in order to have $\tC(\gamma) \ge 0$ we introduced a minus sign in this definition, hence increasing values of $\tC$ corresponds to smaller correlations in the original order parameter $C$. Finally we assume that $\tC(\gamma)$ can be computed by reversing the order of the large $n$ and large $k$ limits, in other words we define for $n$ finite $\tC_n(\g)$ as
\beq
\tC_n(\g) = \lim_{k \to \infty}  (1-C_n(\alpha(k,\g),k))k \ln k \ , 
\label{eq_def_tC}
\eeq
and our third hypothesis will consist in computing $\tC(\gamma)$ as the large $n$ limit of $\tC_n(\g)$.

To rephrase these last two hypotheses, we exclude the possibility that on the scale (\ref{eq_scale}) the fixed point order parameter $C(\alpha,k)$ has distinct scalings with $k$ depending on the value of $\g$, and similarly that the finite distance correlation function $C_n(\alpha,k)$ decays with several plateaus at heights each scaling differently with $k$. These hypotheses are reasonable when confronted with the finite $k$ numerical results (for instance those of Fig.~\ref{fig_k5}), and are somehow corroborated by the scheme of proofs of~\cite{Sly08,MoReTe11_recclus}, that rely on the fact that once $C_n$ is ``small enough'', then it must decay to 0; we shall see in the rest of the paper that they are at least self-consistent. Note also that they hint at the intrisic difficulty of the computation: in the intermediate regime $[\ad(k),\ar(k)]$ there are no hard fields in the fixed point distribution $P(h)$, nevertheless the soft fields become asymptotically hard in the $k \to \infty$ limit as the correlation function $C$ tends to 1 according to (\ref{eq_def_tC_first}).

\subsection{Evolution of the soft fields distribution}

In order to complete the computation of the reduced order parameter we need now to study the large $k$ limit of the distributions $Q_n(h)$ of the soft fields, that obeys the recursion equations (\ref{eq_recursion_Qn},\ref{eq_recursion_hatQn}). Consider first the latter equation; the integer $p$ that appears there is a random number drawn from the binomial distribution ${\rm Bin}(k-1,1-w_n)$. In the large $k$ limit, because of the scaling (\ref{eq_hard_largek}) of $w_n$, this distribution can be approximated by a Poisson distribution ${\rm Po}(x_n/\ln k)$, which hence concentrates on the smallest possible value that appears in the sum, namely $p=1$. We thus obtain at the leading non-trivial order
 \beq
\hQ_n(u) = \left[1 - \frac{3 x_n}{2^{k-1} \ln k} \right] \delta(u) + \frac{3 x_n}{2^{k-1} \ln k} \hR_n(u) \ ,
\label{eq_hQn_largek} 
\eeq
 where $\hR_n(u)$ is a normalized distribution arising from the three contributions of the terms with $p=1$ in (\ref{eq_recursion_hatQn}),
 \beq
\hR_n(u) = \frac{1}{3} \int \dd Q_n(h) \left[ \delta(u-g_1(h)) +  \delta(u-g_1(-h)) + \delta(u+g_1(-h)) \right] \ ,
 \label{eq_hRn_largek} 
\eeq
 where $g_1(h)$ is obtained from (\ref{eq_gp}) as
 \beq
g_1(h) = \frac{\frac{1+h}{2}}{2-\frac{1+h}{2}} \ .
\label{eq_def_g1}
\eeq
Note that this function fulfills the identity $\frac{1-g_1(h)}{1+g_1(h)} = \frac{1-h}{2}$; as a consequence one can check that the Bayes symmetry is respected by $\hR_n$, i.e. that $\hR_n(-u)=\frac{1-u}{1+u} \hR_n(u)$, if $Q_n$ verifies this symmetry.

Let us give an intuitive explanation of the equations (\ref{eq_hQn_largek},\ref{eq_hRn_largek}): in the large $k$ limit the weight of the hard fields $w_n$ is very close to 1, hence almost all of the $k-1$ fields $h_i$ acting on the variables of an hyperedge are forcing them. If they all force them to take the same color $-1$ then to satisfy the constraint the root must take the opposite color, which leads to an hard field $u=1$ (the $k-1$ fields cannot be all forcing in the direction $+1$, this would force the root to be $-1$, in contradiction with the broadcast being conditional on the root equal to $+1$). If they are all hard but with the two colors represented, then the clause is satisfied for every choice of the root, and this produces a trivial soft field $u=0$. The least unprobable situation that can produce a non-trivial soft field ($u\neq 0,1$) is thus when among the $k-1$ entering fields $h_i$ exactly one, say $h_1$, is soft, and all the $k-2$ other are hard. If the two colors appear among the $k-2$ hard fields, again the root is unbiased, this produces a trivial soft field $u=0$. Assume thus that the $k-2$ hard fields are of the same color. If this color is $+1$, then necessarily $\s_1=-1$ and this yields $u=-g_1(-h_1)$. If this color is $-1$, then both $\s_1=+1$ (producing $u=g_1(-h_1)$) and $\s_1=-1$ (corresponding to $u=g_1(h_1)$) are possible; this concludes the interpretation of (\ref{eq_hQn_largek},\ref{eq_hRn_largek}).

We can now inject the simplified form (\ref{eq_hQn_largek}) of $\hQ_n(u)$ in the other recursion equation (\ref{eq_recursion_Qn}); as $f(u_1,\dots,u_l,0,\dots,0)=f(u_1,\dots,u_l)$ the first term of $\hQ_n(u)$ does not contribute, and the number of $\hR_n(u_i)$ picked up in the r.h.s. of  (\ref{eq_recursion_Qn}) is a Poisson random variable of parameter $\alpha k (1-\hw_n) \frac{3 x_n}{2^{k-1} \ln k}$. This quantity converges to $3 x_n$ in the large $k$ limit for the regime we are considering, hence (\ref{eq_recursion_Qn}) becomes:
 \beq
 Q_{n+1}(h) = \sum_{l=0}^\infty e^{-3 x_n} \frac{(3 x_n)^l}{l!} \int \prod_{i=1}^l \dd \hR_n(u_i) \, \delta(h-f(u_1,\dots,u_l)) \ .
 \eeq

 We can finally express the reduced order parameter $C_n(\gamma)$ in the large $k$ limit: combining its definition from (\ref{eq_def_tC}), the expression of $C_n$ in terms of the soft fields distribution and of the hard fields weight given in Eq.~(\ref{eq_Cn_decomposed}), and the scaling of the latter quantity written in (\ref{eq_hard_largek}), we obtain $\tC_n(\g) = x_n \int \dd Q_n(h) (1-h)$.
 
\section{The limit of large distance $n$}
\label{sec_largen}

Let us summarize what we have achieved up to now; starting from the recursion equations (\ref{1RSBeq1},\ref{1RSBeq2}) for the distributions $P_n(h)$ we have taken the large $k$ limit with the density of constraints scaling as $\alpha(k,\g)$ of Eq.~(\ref{eq_scale}), and argued for the existence of a reduced correlation function at finite distance, $\tC_n(\g)$ defined in (\ref{eq_def_tC}). We have then shown that the latter could be computed by solving some recursion equations on both $x_n$, a scalar quantity related to the weight of the hard fields in Eq.~(\ref{eq_hard_largek}), and on $Q_n(h)$, the distribution of the soft fields (that is a distribution supported on $[-1,1]$, with no atoms in $\pm 1$). For the sake of readability we regroup now all the equations that will be necessary to proceed:
\begin{align}
x_0 &= 0 \ , \qquad x_{n+1} = e^{- \g + x_n} \ , \label{eq_summary_xn} \\
Q_1(h)&=\delta(h) \ , \quad Q_{n+1}(h) = \sum_{l=0}^\infty e^{-3 x_n} \frac{(3 x_n)^l}{l!} \int \prod_{i=1}^l \dd \hR_n(u_i) \, \delta(h-f(u_1,\dots,u_l)) \ , \label{eq_summary_Qn}\\
& \qquad f(u_1,\dots,u_l) = \frac{\underset{i=1}{\overset{l}{\prod}} \frac{1+u_i}{2} - \underset{i=1}{\overset{l}{\prod}} \frac{1-u_i}{2} }{\underset{i=1}{\overset{l}{\prod}}  \frac{1+u_i}{2} + \underset{i=1}{\overset{l}{\prod}} \frac{1-u_i}{2}} \ ,\\ 
\hR_n(u) &= \frac{1}{3} \int \dd Q_n(h) \left[ \delta(u-g_1(h)) +  \delta(u-g_1(-h)) + \delta(u+g_1(-h)) \right] \ , \qquad g_1(h) = \frac{\frac{1+h}{2}}{2-\frac{1+h}{2}} \ , \label{eq_summary_hRn} \\
\tC_n(\g) &= x_n \int \dd Q_n(h) (1-h) \label{eq_summary_tCn} \ .
 \end{align}
Let us emphasize that the parameter $k$ has disappeared from these equations, that only depend on $\g$; moreover we show in the Appendix that exactly the same equations describe the large $q$ limit of the coloring problem. What remains to do now is to study the large $n$ limit of these equations, as the threshold $\gd$ will be determined according to the behavior of $\tC_n(\gamma)$ in this limit, namely $\tC_n(\gamma) \to +\infty$ will signal that $\g < \gd$, while $\tC_n(\gamma)$ will remain finite for $n\to\infty$ if $\g > \gd$.

\subsection{The regime $\g \ge \gr = 1$}
\label{sec_largen_largeg}

\begin{figure}
\begin{center}
 \includegraphics[width=8cm]{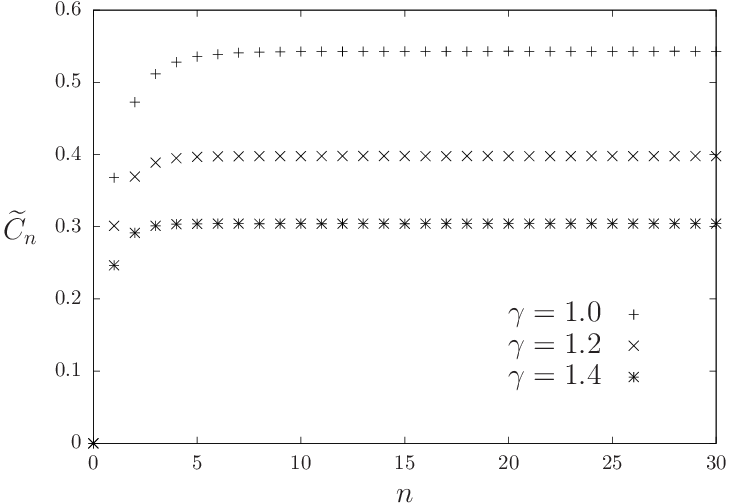}
 \hspace{1cm}
  \includegraphics[width=8cm]{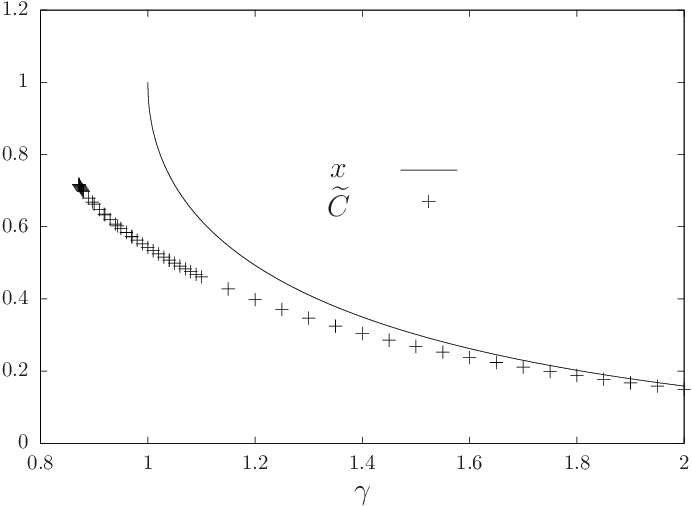}
  \caption{Left: the reduced correlation function $\tC_n(\g)$ as a function of $n$ for a few values of $\g \ge \gr =1$.
    Right: the large $n$ limit $\tC(\g)=\lim_n \tC_n(\g)$ as a function of $\g$, along with the contribution of the hard fields $x(\g)=\lim_n x_n(\g)$. The first curve exists for $\g \ge \gd \approx 0.871$, the second one for $\g \ge \gr=1$. The data for $\g <1$ have been obtained through the reweighted algorithm expained in Sec.~\ref{sec_largen_reweighted}.}
\label{fig_largek}
\end{center}
\end{figure}

Let us start by considering the case $\g \ge \gr = 1$; then the sequence $x_n$ remains bounded for all $n$ (its limit $x(\g)$ being depicted as a function of $\g$ on the right panel of Fig.~\ref{fig_largek}), which translates the presence of hard fields in the fixed point solution of the corresponding finite $k$ regime $\alpha \ge \ar$. As a consequence the numerical resolution of (\ref{eq_summary_Qn}) presents no difficulty and can be performed with the usual population dynamics algorithms explained in Sec.~\ref{sec_num_finitek}, $Q_n(h)$ being represented as an empirical distribution over a sample of $h_i$'s. The figure \ref{fig_largek} presents some numerical results obtained in this way; one can see on the left panel that $\tC_n(\g)$ converges to a finite limit as $n$ grows, this limit value being drawn as a function of $\g$ on the right panel. In this regime $Q_n$ converges (weakly) to a stationary distribution $Q$, solution of the fixed point equation obtained by replacing in (\ref{eq_summary_Qn}) $x_n$ by its limit.

The finite $k$ inequalities $w_n \le C_n \le 1$ becomes in these rescaled units $0 \le \tC_n(\g) \le x_n(\g)$; indeed,
\beq
\int \dd Q_n(h) (1-h) = 1- \int \dd Q_n(h) h = 1- \int \dd Q_n(h) h^2 \in [0,1] \ ,
\eeq
because of the consequence stated in (\ref{eq_Bayes_moment}) of the Bayes symmetry enjoyed by $Q_n$. In the large $n$ limit we thus have $\tC(\g) \le x(\g)$, as illustrated in the right panel of Fig.~\ref{fig_largek} which is the analog of the right panel of Fig.~\ref{fig_k5} for finite $k$.

Note that $\tC(\g)$ seems to behave smoothly as $\g \to \gr^+$, as expected if the transition occurs at $\gd < \gr$. As $x$ has a square root singularity this means that there is also a square root singularity in $\int \dd Q(h) (1-h)$ as a function of $\gamma$, in order for their product $\tC(\g)$ to be regular.

\subsection{The difficulties for $\g < 1$}

\begin{figure}
\begin{center}
\includegraphics[width=8cm]{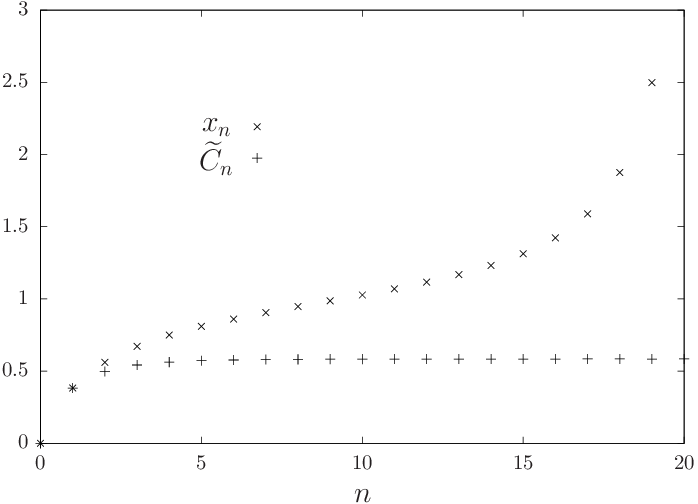}
\caption{The evolution of $\tC_n$ and $x_n$ for $\g =0.96 < \gr$. The sequence $x_n$ diverges asymptotically in a very rapid way (as iterated exponentials, the vertical range has been cut for the sake of readability) while the reduced correlation function $\tC_n$ tends to a positive constant. This is the analog of the finite $k$ regime $\ad < \alpha <\ar$ depicted in Fig.~\ref{fig_k5_a10.40}.}
\label{fig_g0.96}
\end{center}
\end{figure}

The study of the large $n$ limit is much more difficult when $\g < 1$: the sequence $x_n$ solution of (\ref{eq_summary_xn}) is then divergent (asymptotically very strongly, as iterated exponentials). As the direct implementation of the population dynamics procedure to solve (\ref{eq_summary_Qn}) involves the manipulation of a number of fields $u_i$'s that is distributed as a Poisson random variable of parameter $x_n$, this becomes very quickly impossible to handle. A first hint of what happens for $\g < 1$ can nevertheless be unveiled by this numerical procedure: when $\g$ is only slightly smaller than $1$, $x_n$ grows rather slowly at first, the range of $n$ that can be treated in this way is then sufficiently large to gain some useful informations. A typical result obtained in this way is plotted on Fig.~\ref{fig_g0.96} for $\g=0.96$: even though $x_n$ is diverging it is tempting to conjecture from this plot that $\tC_n$ admits a limit when $n \to \infty$. Looking at the expression (\ref{eq_summary_tCn}) of $\tC_n$ one realizes that this is possible provided $\int \dd Q_n(h)(1-h)$ tends to $0$ as $n$ grows, with an order of magnitude inversely proportional to the one of $x_n$. This weak convergence of $Q_n(h)$ to $\delta(h-1)$ is compatible with the form of the recursion equation (\ref{eq_summary_Qn}), the combination of a very large number of $u_i$'s can indeed produce a field $h$ very close to 1.

Lowering further the value of $\g$ the speed of the divergence of $x_n$ increases so much that an extrapolation of $\tC_n$ to large $n$ values becomes impossible with this numerical method, which as a consequence does not allow for an accurate estimation of $\gd$. We thus need to devise a formalism that incorporates the compensation mechanism between the divergence of $x_n$ and the concentration of $Q_n$ around $\delta(h-1)$ at the origin of the finite value of $\tC$. A natural idea would be to make the change of variable $h=1-(h'/x_n)$, with $h'$ a random variable that should remain finite and thus obey a simplified recursion equation. Unfortunately the Bayes symmetry enjoyed by $Q_n$, and its consequences on the moments of $h$, imply that $h'$ has higher moments that diverge with $n$. This strategy seems thus rather unpractical. In fact it is more probable that the value of $\tC_n$ is determined by contributions of all values of $h$ in $(-1,1)$, even very improbable ones where $Q_n$ is of order $1/x_n$, while the typical values around $h=1$ do not contribute to $\tC_n$ because of the factor $1-h$ in the integral in (\ref{eq_summary_tCn}). In more technical terms we conjecture that $x_n Q_n$, which is a sequence of positive measures, even if not probability ones because their total mass is the diverging sequence $x_n$, does converge to a positive measure $\rho$, in the sense of the vague convergence on $(-1,1)$, i.e.
\beq
\lim_{n \to \infty }x_n \int_a^b \dd Q_n(h) = \int_a^b \dd \rho(h)    
\eeq
for all $-1<a<b<1$ which are continuity points of $\rho$. A direct numerical test of this conjecture, and the determination of $\rho$, seems rather difficult as $\rho$ is not a probability measure and may have an infinite total mass. We have thus devised an alternative numerical procedure that allowed us to explore the regime $\g<1$, as we shall now explain.

\subsection{Reweighted probability distributions}
\label{sec_largen_reweighted}

The idea underlying this procedure is to study the evolution not of the probability measure $Q_n$, but of a reweighted (or tilted) version of it, that give less importance to the typical (for $Q_n$) values of $h$ around $h=1$, which do not contribute to $\tC_n$. Let us thus consider a positive function $r_n(h)$, and define the reweighted distribution $\mu_n(h) = x_n r(h) Q_n(h)$. In more precise mathematical terms $\mu_n$ is absolutely continuous with respect to $Q_n$, with relative density $x_n r$. For the reasons explained above our goal is to use a function $r(h)$ that vanishes in $h=1$; a convenient choice is to take $r$ of the form $r(h)=((1-h)/(1+h))^t$, with $t$ an exponent that is arbitrary for the moment. Indeed with this choice the reweighting will factorize in (\ref{eq_summary_Qn}), in the sense that $r(f(u_1,\dots,u_l))=r(u_1)\times \dots \times r(u_l)$. We will actually further specify the function by taking the exponent $t=1/2$, thus setting
\beq
\mu_n(h) = x_n  Q_n(h) \left(\frac{1-h}{1+h}\right)^{\frac{1}{2}} \ .
\label{eq_def_mun}
\eeq
The reasons for this choice of the exponent are two fold: the Bayes symmetry (\ref{eq_Bayes}) enjoyed by $Q_n$ implies then that $\mu_n$ is symmetric in the usual sense, i.e. $\mu_n(-h)=\mu_n(h)$. Furthermore (\ref{eq_Bayes}) also implies the following identity (used in the proof of lemma 4.4 in~\cite{MoReTe11_recclus}), 
\beq
\int \dd Q_n(h) \left(\frac{1-h}{1+h} \right)^{\frac{1}{2}} = \int \dd Q_n(h) \sqrt{1-h^2} \ ,
\label{eq_property_mass}
\eeq
which can be obtained from (\ref{eq_Bayes}) with the choice $b(h)=h((1-h)/(1+h))^{1/2}$. The right hand side of (\ref{eq_property_mass}) is obviously between 0 and 1, thus for all finite $n$ the measure $\mu_n$ has a finite total mass, smaller than $x_n$. We shall therefore define a probability measure $\nu_n$ by dividing $\mu_n$ by its mass $m_n$:
\beq
m_n = \int \dd \mu_n(h) =  x_n \int \dd Q_n(h) \left(\frac{1-h}{1+h} \right)^{\frac{1}{2}} \ , \qquad \nu_n(h) = \frac{1}{m_n} \mu_n(h) \ .
\eeq
Note that the three descriptions in terms of the (finite but not normalized to 1) positive measure $\mu_n$, or the pair $(m_n,\nu_n)$, or the original one in terms of $(x_n,Q_n)$, are strictly equivalent for all finite $n$; in particular one can compute the reduced correlation function as
\beq
\tC_n(\g) = \int \dd \mu_n(h) \sqrt{1-h^2} = m_n \int \dd \nu_n(h) \sqrt{1-h^2} \ . 
\label{eq_tCn_reweighted}
\eeq
Besides its equivalence with the original quantities at finite $n$ the advantage of the reweighted description in terms of $\mu_n$ is a much nicer behavior when $n$ grows, as we shall see.

Let us now derive the recursion equations that governs the evolution of $\mu_n$ (or equivalently of the pair $(m_n,\nu_n)$). It will be more convenient to express some quantities in terms of $\tih = \text{arcth}(h)$ and $\tiu = \text{arcth}(u)$, that corresponds to the effective magnetic fields associated to the magnetizations $h$ and $u$. In particular the operation $h=f(u_1,\dots,u_l)$ translates in this domain to a simple addition, $\tih=\tiu_1 + \dots + \tiu_l$, and the reweighting $r(h)$ becomes $e^{-\tih}$. For the sake of conciseness we shall take the liberty of using simultaneously both types of quantities in the following equations, keeping implicit the relation $\tih = \text{arcth}(h)$ between them; we will also denote with the same symbol $Q_n$ the measure for the random variables $h$ and $\tih$ related in this way.

Let us first rewrite the equations (\ref{eq_summary_Qn},\ref{eq_summary_hRn}) in terms of $\tih$:
\begin{align}
Q_{n+1}(\tih) &= \sum_{l=0}^\infty e^{-3 x_n} \frac{(3 x_n)^l}{l!} \int \prod_{i=1}^l \dd \hR_n(\tiu_i) \, \delta(\tih-\tiu_1-\dots-\tiu_l)) \ , \label{eq_Qn_tih} \\
\hR_n(\tiu) &= \frac{1}{3} \int \dd Q_n(\tih) \left[ \delta(\tiu-a(\tih)) +  \delta(\tiu-a(-\tih)) + \delta(\tiu+a(-\tih)) \right] \ , \label{eq_hRn_tiu}
\end{align}
where we introduced the function
\beq
a(\tih) = \text{arcth}(g_1(\tanh(\tih))) = \frac{1}{2} \ln\left(1+e^{2 \tih} \right) \ .
\label{eq_def_a}
\eeq
Under $Q_{n+1}$ the random variable $\tih$ has a compound Poisson distribution, it is thus more convenient to describe it in terms of its characteristic function (i.e. Fourier transform). We define the latter for the various measures as 
\beq
\hQ_n(z)= \int \dd Q_n(h) e^{i z \tih} \ , \qquad
\hmu_n(z) = \int \dd \mu_n(h) e^{i z \tih}  \ , \qquad
\hnu_n(z) = \int \dd \nu_n(h) e^{i z \tih} = \frac{1}{m_n} \hmu_n(z) \ .
\eeq
With these conventions the two equations (\ref{eq_Qn_tih},\ref{eq_hRn_tiu}) become:
\beq
\hQ_{n+1}(z)=\exp\left[-3 x_n + x_n \int \dd Q_n(h) \left(
e^{i z a(\tih)} + e^{i z a(-\tih)} + e^{-i z a(-\tih)} 
  \right)\right] \ .
\eeq
The effect of the reweighting is easily seen to translate in terms of the characteristic functions as a shift of the argument,
\beq
\mu_{n+1}(h) = x_{n+1} e^{- \tih} Q_{n+1}(h) \quad \Leftrightarrow \quad
\hmu_{n+1}(z) = x_{n+1} \hQ_{n+1}(z+i) \ .
\eeq
Recalling that $x_{n+1}=\exp[-\g +x_n]$ we thus obtain
\beq
\hmu_{n+1}(z) = \exp\left[-\g + x_n \int \dd Q_n(h) \left( e^{- a(\tih)} e^{i z a(\tih)} + 
 e^{- a(-\tih)} e^{i z a(-\tih)} + e^{ a(-\tih)} e^{-i z a(-\tih)} -2
  \right)\right]
\eeq
Replacing in this equation $x_nQ_n$ by $e^{\tih} \mu_n$ yields
\begin{align}
  \hmu_{n+1}(z) &= \exp\left[-\g + \int \dd \mu_n(h) \left( e^{\tih - a(\tih)} e^{i z a(\tih)}
+  e^{\tih - a(-\tih)} e^{i z a(-\tih)} + e^{\tih+ a(-\tih)} e^{-i z a(-\tih)} - 2 e^{\tih} \right)\right] \\
&= \exp\left[-\g + \int \dd \mu_n(h) \left(
\sqrt{\frac{1+h}{2}} e^{i z a(\tih)}
  + \sqrt{\frac{1+h}{1-h}} \left(\sqrt{\frac{1+h}{2}}    e^{i z a(-\tih)}
  + \sqrt{\frac{2}{1+h}} e^{-i z a(-\tih)}
  - 2 
\right) \right)\right]
\end{align}
where in the second line we expressed the weight factors in terms of $h$ instead of $\tih$, using the expression of $a(\tih)$ given in (\ref{eq_def_a}). As $\mu_n(h)=\mu_n(-h)$ one can symmetrize the integrand to obtain
\begin{align}
\hmu_{n+1}(z) &= \exp\left[-\g + \int \dd \mu_n(h) \left(
\frac{1}{\sqrt{2(1+h)}} \left( e^{i z a(\tih)} + e^{-i z a(\tih)} \right)
+ \frac{1}{\sqrt{2(1-h)}} \left( e^{i z a(-\tih)} + e^{-i z a(-\tih)} \right)
- \frac{2}{\sqrt{1-h^2}}
                \right)\right]  \nonumber \\
  &= \exp\left[-\g + \int \dd \mu_n(h) \left(
 \sqrt{\frac{2}{1-h}} \left( e^{i z a(-\tih)} + e^{-i z a(-\tih)} \right)
- \frac{2}{\sqrt{1-h^2}}
    \right)\right] \ .
    \label{eq_hmunp1}
\end{align}
These two expressions for $\hmu_{n+1}(z)$ are obviously invariant under $z \to -z$, which enforces the conservation of the symmetry $\mu_{n+1}(h)=\mu_{n+1}(-h)$ along the iterations. The integrand of the first one is explicitly symmetric in $h$, while the more compact second one has been obtained by exploiting the symmetry of $\mu_n$.

We have thus obtained a recursion equation for the reweighted measure $\mu_n$, that is equivalent to the original system on $(x_n,Q_n)$. It will be useful to spell out its translation in terms of the mass $m_n$ of $\mu_n$ and its normalized version $\nu_n$. By setting $z=0$ in (\ref{eq_hmunp1}) we obtain
\beq
m_{n+1} = \exp\left[-\g + \int \dd \mu_n(h) \beta(h) \right] \ , \qquad
\text{with} \ \ \beta(h) = \sqrt{\frac{2}{1+h}} + \sqrt{\frac{2}{1-h}} - \frac{2}{\sqrt{1-h^2}} \ .
\label{eq_mnp1}
\eeq
Note that $\beta(h) \le 1$ for $h \in [-1,1]$, hence the integral in (\ref{eq_mnp1}) is smaller than $m_n$, which allows to check inductively that $m_n \le x_n$, in agreement with the direct derivation of this bound from (\ref{eq_property_mass}). Dividing the expression (\ref{eq_hmunp1}) of $\hmu_{n+1}(z)$ by its total mass $m_{n+1}$ we obtain the characteristic function $\hnu_{n+1}(z)$ of $\nu_{n+1}$, the normalized probability distribution. Replacing $\mu_n$ by $m_n \nu_n$ we thus have a recursion equation on the pair $(m_n,\nu_n)$, namely
\begin{align}
m_{n+1} &= \exp\left[-\g + m_n \int \dd \nu_n(h) \beta(h) \right] \ , \label{eq_mnp1_nu} \\
\hnu_{n+1}(z) &=\exp \left[m_n\int \dd \nu_n(h) \sqrt{\frac{2}{1-h}} \left(e^{i z a(-\tih)}   +  e^{-i z a(-\tih)}  - 2 \right) \right] \ , \label{eq_hnu_nu}
\end{align}
completed by the initial condition $m_1=e^{-\g}$, $\nu_1(h)=\delta(h)$, that derives directly from $x_1=e^{-\g}$, $Q_1(h)=\delta(h)$.

\subsection{Numerical resolution}
\label{sec_largen_numerical}

There is one last difficulty to face in the derivation of a numerical procedure able to solve the equations (\ref{eq_mnp1_nu},\ref{eq_hnu_nu}), that we shall now explain. As $\nu_n$ is a normalized probability measure we can represent it as a large sample of fields, according to the population dynamics strategy explained in Sec.~\ref{sec_num_finitek}. Suppose this has been done up to the $n$-th iteration, and that an estimate of the real number $m_n$ is also known. Then $m_{n+1}$ can be easily computed from (\ref{eq_mnp1_nu}), the integration over $\nu_n$ being estimated as an empirical average over the representative population. There remains to give an interpretation of the law $\nu_{n+1}$ whose characteristic function is given in (\ref{eq_hnu_nu}), and to devise a sampling procedure from it in order to produce the population of i.i.d. samples that will represent $\nu_{n+1}$. In order to do so let us introduce a positive (not normalized) measure $\pi_n$ and its total mass $\lambda_n$ according to
\beq
\pi_n(y) =  m_n  \int \dd \nu_n(h)  \sqrt{\frac{2}{1-h}}(\delta(y-a(-\tih)) + \delta(y+a(-\tih))) \ , \qquad
\lambda_n = 2 m_n \int \dd \nu_n(h)  \sqrt{\frac{2}{1-h}} 
 \ ,
\label{eq_def_pi_lambda}
 \eeq
in such a way that (\ref{eq_hnu_nu}) can be rewritten
\beq
\hnu_{n+1}(z) =\exp \left[ \int \dd \pi_n(y) (e^{i z y}-1) \right] \ .
\eeq
In more technical terms $\pi_n$ is the L\'evy measure associated to the infinitely divisible distribution $\nu_{n+1}$~\cite{book_Sato}, with $\lambda_n$ the total mass of $\pi_n$, which is finite for a Poisson compound distribution. This form makes clear that under $\nu_{n+1}$ the random variable $\tih$ can be described as the distributional equality $\tih \eqd \sum_{i=1}^l y_i$, where in the right hand side $l$ is a Poisson random variable of mean $\lambda_n$, and the $y_i$'s are i.i.d. copies drawn with the probability measure $\pi_n(y)/\lambda_n$. If $\nu_n$ is known as a sample of fields it is possible to sample efficiently $y$ from $\pi_n/\lambda_n$, by extracting a field $h_i^{(n)}$ with a probability proportional to $(1-h_i^{(n)})^{-1/2}$ and then setting $y = \pm a(-\tih_i^{(n)})$, the two signs being chosen with probability $1/2$. It could thus seem at first sight that the distributional interpretation given above leads to a valid numerical procedure to solve (\ref{eq_mnp1_nu},\ref{eq_hnu_nu}). A closer look reveals that this strategy cannot work as it is for $\g < 1$. Let us indeed rewrite the expression of $\lambda_n$ by exploiting the symmetry of $\nu_n$:
\beq
\lambda_n = m_n \int \dd \nu_n(h) \left( \sqrt{\frac{2}{1-h}} + \sqrt{\frac{2}{1+h}} \right) = m_n \int \dd \nu_n(h) \frac{\sqrt{2}(\sqrt{1+h} +\sqrt{1-h} ) }{\sqrt{1-h^2}} \ .
\eeq
On the other hand the sequence $x_n$ can be expressed from the definition (\ref{eq_def_mun}) of $\mu_n$ as 
\beq
x_n = \int \dd \mu_n(h) \sqrt{\frac{1+h}{1-h}} = m_n \int \dd \nu_n(h) \frac{1}{2} \left( \sqrt{\frac{1+h}{1-h}}  + \sqrt{\frac{1-h}{1+h}} \right)
= m_n \int \dd \nu_n(h) \frac{1}{\sqrt{1-h^2}} \ ,
\eeq
where we used the normalization of $Q_n$ and the symmetry of $\nu_n$. Noting that the function $\sqrt{2}(\sqrt{1+h} +\sqrt{1-h} )$ is bounded between $2$ and $2 \sqrt{2}$ when $h \in [-1,1]$ we can conclude by comparing these two expressions of $\lambda_n$ and $x_n$ that $2 x_n \le \lambda_n \le 2 \sqrt{2} x_n$. Hence for $\g < 1$ the lowerbound implies that $\lambda_n$ diverges with $n$, which brings us back to the situation we are trying to avoid of a Poisson compound distribution with a diverging number of summands.

Fortunately the reweighting performed above will allow us to circumvent this difficulty. One sees indeed on the expression (\ref{eq_def_pi_lambda}) that the divergent contribution to $\lambda_n$ arises from the neighborhood of $h=1$, i.e. of $\tih=+\infty$. However the corresponding summands $\pm a(-\tih)$ are very small (recall the expression of $a$ from Eq.~(\ref{eq_def_a})), it is thus conceivable that such a sum of a very large number of very small contributions can be handled in a simplified way. To formalize this intuition we rewrite (\ref{eq_hnu_nu}) as
\begin{align}
\hnu_{n+1}(z) &= \hnu^{(\le)}_{n+1}(z) \, \hnu^{(>)}_{n+1}(z)  \ , \\
\hnu^{(\le)}_{n+1}(z) &=\exp \left[\int \dd \mu_n(h) \sqrt{\frac{2}{1-h}} \left(e^{i z a(-\tih)}   +  e^{-i z a(-\tih)}  - 2 \right) \ind(h \le 1 -\ve_n) \right] \ , \\
\hnu^{(>)}_{n+1}(z)&=\exp \left[\int \dd \mu_n(h)\sqrt{\frac{2}{1-h}} \left(e^{i z a(-\tih)}   +  e^{-i z a(-\tih)}  - 2 \right)\ind(h > 1 -\ve_n) \right] \ ,
\end{align}
where $\ve_n$ is for the moment an arbitrary threshold; such a decomposition is loosely inspired by an analogy with the L\'evy-It\^o decomposition of L\'evy processes~\cite{book_Sato}.

These two contributions to $\hnu_{n+1}(z)$ are characteristic functions of probability distributions, hence under the law $\nu_{n+1}$ the random variable $\tih$ is the sum of independent draws from $\nu^{(\le)}_{n+1}$ and $\nu^{(>)}_{n+1}$. The first one can be described as above, modulo the introduction of the cutoff $\ve_n$; defining
\begin{align}
\pi_n^{(\le)}(y) &=  m_n  \int \dd \nu_n(h)  \sqrt{\frac{2}{1-h}}(\delta(y-a(-\tih)) + \delta(y+a(-\tih))) \ind(h \le 1 -\ve_n) \ , \\
\lambda_n^{(\le)} &= 2 m_n \int \dd \nu_n(h)  \sqrt{\frac{2}{1-h}} \, \ind(h \le 1 -\ve_n) \ ,
 \end{align}
one sees that under the law $\nu^{(\le)}_{n+1}$ the random variable $\tih$ satisfies the distributional equality
$\tih \eqd \sum_{i=1}^l y_i$, where in the right hand side $l$ is a Poisson random variable of mean $\lambda_n^{(\le)}$ and the $y_i$ are i.i.d. with the law $\pi_n^{(\le)}/\lambda_n^{(\le)}$.

The law of $\nu^{(>)}_{n+1}$ will instead be described via its cumulants, to be denoted $c_{n+1,p}$ for the $p$-th one. By taking derivatives with respect to $z$ of $\ln \hnu^{(>)}_{n+1}(z)$ one finds easily that
\beq
c_{n+1,p} = m_n \int \dd \nu_n(h)\sqrt{\frac{2}{1-h}} (1+(-1)^p)  \, a(-\tih)^p \, \ind(h > 1 -\ve_n) \ .
\eeq
As expected from the symmetry of the law $\nu^{(>)}_{n+1}$ all its cumulants of odd order vanish and one can write for the even ones:
\beq
c_{n+1,2p} = 2 m_n \int \dd \nu_n(h) \sqrt{\frac{2}{1-h}} \, a(-\tih)^{2p} \, \ind(h > 1 -\ve_n) \ .
\eeq

The strategy we have followed in our numerical resolution is to choose $\ve_n$ large enough such that $\lambda_n^{(\le)}$ does not grow with $n$, in such a way that the contribution from $\nu^{(\le)}_{n+1}$ can be generated with a finite Poissonian number of summands, but small enough so that $\nu^{(>)}_{n+1}$ can be safely approximated as a Gaussian distribution, neglecting the cumulants $c_{n+1,2p}$ for $p>1$ (there exist some ways to draw better approximations of infinitely divisible distributions by using more cumulants~\cite{Asmussen01,Chi14}, but we did not try to implement them). Let us give an explicit description of the algorithm. At the $n$-th iteration we assume to have an estimation of $m_n$ and of $\nu_n$ as a population, namely
\beq
\nu_n(h) \approx \frac{1}{\N} \sum_{i=1}^\N \delta(\tih -\tih_i^{(n)}) \ .
\label{eq_nu_population}
\eeq
We will assume that the population has been sorted, $\tih_1^{(n)} \le \tih_2^{(n)} \le \dots \le \tih_{\N}^{(n)} $, and translate the cutoff $\ve_n$ by defining $\N_n$ in such a way that $h_{\N_n}^{(n)} \le 1-\ve_n < h_{\N_n + 1}^{(n)}$. We can thus estimate all integrals with respect to $\nu_n$ according to
\beq
\int \nu_n(h) F(\tih) \approx \frac{1}{\N} \sum_{i=1}^\N F(\tih_i^{(n)})
\label{eq_sampling}
\eeq
for all functions $F$, and in particular this gives us $m_{n+1}$ from (\ref{eq_mnp1_nu}). We then compute
\beq
\lambda_n^{(\le)}  \approx
2 m_n \frac{1}{\N}\sum_{i=1}^{\N_n} \sqrt{\frac{2}{1-h_i^{(n)}}} \ , \qquad
c_{n+1,2} \approx 2 m_n \frac{1}{\N} \sum_{i=\N_n+1}^\N \sqrt{\frac{2}{1-h_i^{(n)}}}  a(-\tih_i^{(n)})^2 \ .
\label{eq_lambda_numeric}
\eeq
We then construct the new population elements $\tih_i^{(n+1)}$ by repeating $\N$ times, independently for $i=1,\dots,\N$:
\begin{itemize}
\item draw a random number $l$ from a Poisson distribution of parameter $\lambda_n^{(\le)}$;
\item draw $i_1,\dots,i_l$ i.i.d. in $\{1,\dots,\N_n\}$ with probability proportional to $\frac{1}{\sqrt{1-h_i^{(n)}}}$ (this can be done efficiently by precomputing a cumulative table);
\item set $\tih_i^{(n+1)} = \delta_1 \, a(-\tih_{i_1}^{(n)})+\dots+ \delta_l \, a(-\tih_{i_l}^{(n)}) + \sqrt{c_{n+1,2}} z$ where the $\delta_i$'s are $\pm 1$ with probability $1/2$, independently from each other and from anything else, and $z$ is a standard Gaussian random variable.
\end{itemize}
We can then compute the reduced correlation function $\tC_{n+1}$ from (\ref{eq_tCn_reweighted}), and sort the elements of the new population to be ready for the next step.

In our implementation we chose the threshold $\ve_n$, or equivalently $\N_n$, in an adaptive way: at each step we take the largest value of $\N_n$ that maintains $\lambda_n^{(\le)}$ computed from (\ref{eq_lambda_numeric}) smaller than a value $\lambda$ fixed beforehand and kept constant along the iterations. The numerical accuracy of our procedure is thus limited by the finiteness of $\N$ (the representation (\ref{eq_nu_population}) becomes exact only in the limit $\N \to \infty$) and of $\lambda$ (because of the Gaussian approximation we perform for $\nu_{n+1}^{(>)}$). The numerical results presented in the figures have been obtained with $\N=10^7$ and $\lambda=20$; we checked that varying $\lambda$ between $10$ and $30$, or reducing $\N$ to $10^6$, did not affect our conclusions in a quantitative way.

\begin{figure}
\begin{center}
  \includegraphics[width=8cm]{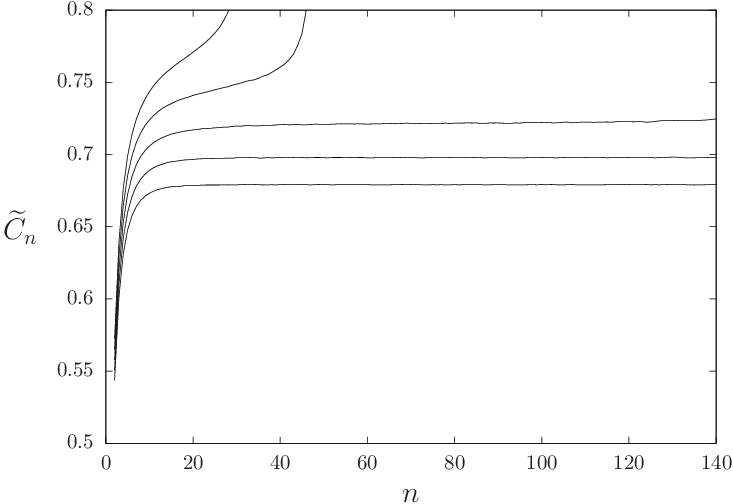}
  \hspace{1cm}
\includegraphics[width=8cm]{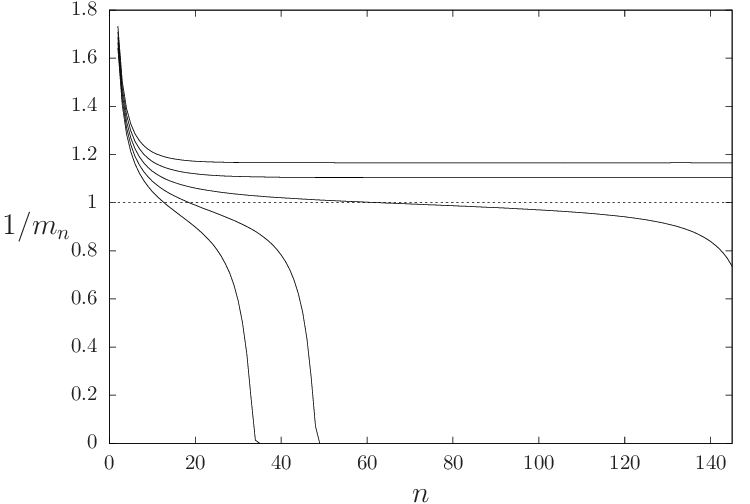}
\caption{The results of the reweighted distribution method in the neighborhood of $\gd$: on the left panel the reduced order parameter $\tC_n(\g)$ as a function of $n$, on the right panel the inverse of the mass $m_n$ of the reweighted measure $\mu_n$. On both panels the five curves correspond to $\g=0.85$, $\g=0.86$, $\g=0.87$, $\g=0.88$, $\g=0.89$, from top to bottom on the left panel and from bottom to top on the right panel. These curves show that $0.87<\gd<0.88$, $\tC_n$ and $m_n$ diverge for $\g=0.87$, even if it is not completely visible on the displayed range of $n$.}
\label{fig_mn}
\end{center}
\end{figure}

On the left panel of Fig.~\ref{fig_mn} we show that this new method allows us to track the evolution of $\tC_n$ for values of $\g$ deep in the regime $\g <1$ that was unaccessible to the simplest strategy presented in Sec.~\ref{sec_largen_largeg}. In particular for $\g \gtrsim 0.88$ there is a clear plateau in $\tC_n$, the value of which was reported as a function of $\g$ in the right panel of Fig.~\ref{fig_largek} along with the results previously obtained for $\g \ge 1$. Note that the new algorithm is also valid for $\g >1$ (and in this case the cutoff $\ve_n$ can be safely put to 0), we found as expected a perfect agreement between the results of the two procedures in this case.

On the right panel of Fig.~\ref{fig_mn} we have plotted the evolution of (the inverse of) $m_n$ as a function of $n$, that presents the same kind of plateau behavior as $\tC_n$. These numerical results suggest that $\tC_n$ diverges with $n$ if and only if $m_n$ diverges, and hence that $\gd$ can be defined as the smallest value of $\g$ such that $m_n$ remains bounded when $n \to \infty$. We have not been able to prove analytically this statement, that will be taken as an additional hypothesis in the rest of the paper. This conjecture could be wrong if there existed a regime of $\g$ such that $\tC_n$ remains finite while $m_n$ diverges, because of the existence of another reweighting scheme that deals with the compensation between the divergence of $x_n$ and the convergence of $Q_n(h)$ to $\delta(h-1)$ in a more efficient way than the one we introduced in (\ref{eq_def_mun}). We found numerically no trace of such a phenomenon, which if it existed would only modify our estimate of $\gd$ by a quantitatively very small amount, as for $\g=0.87$ the finite $n$ study strongly suggests a common divergence of $\tC_n$ and $m_n$ (and for $\g=0.86$ the divergences are obvious on Fig.~\ref{fig_mn}).

\subsection{The fixed point equation and the determination of $\gd$}

\begin{figure}
\begin{center}
  \includegraphics[width=8cm]{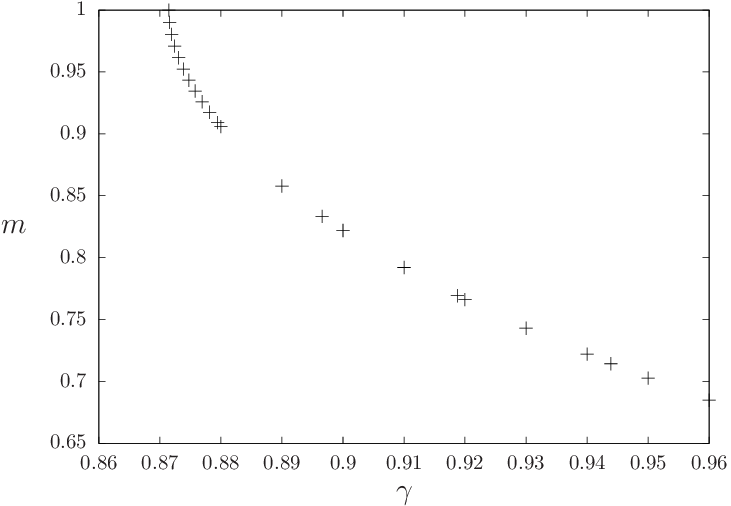}
  \caption{The relationship between the parameter $\g$ and the mass $m$ of the reweighted measure $\mu$ in the $n \to \infty$ limit. The numerical results suggest that $m$ reaches 1 with a square root behavior as $\g$ tends to $\gd$.}
    \label{fig_mofgamma}
\end{center}
\end{figure}

The numerical results presented above suggest that for $\g \ge \gd$ both $m_n$ and $\tC_n$ converge when $n \to \infty$. It is then tempting to conjecture that the probability distribution $\nu_n$ itself converges to some $\nu$, in other words that we can remove the indices $n$ from (\ref{eq_mnp1_nu},\ref{eq_hnu_nu}) and get a system of self-consistent equations on the fixed-point $(m,\nu)$ as:
\begin{align}
m &= \exp\left[-\g + m\int \dd \nu(h) \beta(h) \right] \ , \label{eq_fp_m} \\
\hnu(z) &=\exp \left[ \int \dd \pi(y) (e^{i z y}-1) \right] \ , \qquad
\pi(y) =  m  \int \dd \nu(h)  \sqrt{\frac{2}{1-h}}(\delta(y-a(-\tih)) + \delta(y+a(-\tih))) \ .
\label{eq_fp_nu}
\end{align}
Let us make a few remarks on these equations; first, one can check that if $m$ is finite and $\nu$ is a probability measure with no atoms in $h=\pm 1$ (i.e. in $\tih = \pm \infty$), then the measure $\pi$ defined in (\ref{eq_fp_nu}) gives a finite mass to all sets that are bounded away from $0$, does not have an atom in $0$, and is such that $\int \dd \pi(y) |y| \ind(|y| \le 1) < \infty$. This ensures that the integral in the definition of $\hnu(z)$ converges, hence $\pi$ is a valid L\'evy measure for the infinitely divisible distribution $\nu$. Moreover the assumption of convergence of $\nu_n$ towards $\nu$ is self-consistent in the following sense: if $\nu_n \tow \nu$ (i.e. in the sense of weak convergence of probability measures) then $\pi_n(f) \to \pi(f)$ for all bounded continuous functions $f$ that vanish in a neighborhood of $0$, and $\lim_{\epsilon \to 0} \limsup_{n\to \infty} \int \dd \pi_n(y) y^2 \ind(|y| \le \epsilon) =0 $. These two properties allow to invoke theorem 8.7 of \cite{book_Sato} to conclude that $\nu_n \tow \nu$, the infinitely divisible distribution with L\'evy measure $\pi$. Of course this circular argument does not prove the convergence $\nu_n \tow \nu$, but at least shows its consistency. 

Note that in (\ref{eq_fp_m},\ref{eq_fp_nu}) we have finally achieved the main goal of the article, namely getting rid of the parameters $k$ and $n$ that have been sent to infinity and obtaining a system of equations that only depend on the parameter $\g$. Moreover $\g$ only appears in (\ref{eq_fp_m}): one can thus view (\ref{eq_fp_nu}) as a self-consistent equation on $\nu$, parametrized by $m$, from which one deduces $\g$ according to (\ref{eq_fp_m}). We have performed this numerical resolution at fixed $m$, relying again on the Gaussian approximation to deal with the divergence of the total mass of $\pi$, and checked that the results were in perfect agreement with the one obtained fixing $\g$ in the $n \to \infty$ limit. In Fig.~\ref{fig_mofgamma} we have plotted the correspondance function between $\g$ and $m$. The numerical resolution of the fixed point equation (\ref{eq_fp_nu}) led us to the following observations: (i) a solution of the equation only exists for $m \le 1$ (ii) the tail behavior of $\nu$ is very well described numerically by
\beq
\nu(\{\tih \ge x \} ) \sim S(x) \frac{e^{-r(m) x}}{x^\rho} \qquad \text{as} \ \ x \to +\infty \ ,
\label{eq_tail_nu}
\eeq
where $S$ is a slowly varying function, $\rho$ is an exponent very close to $1$ for all the values of $m$ we have considered, and $r(m) \ge 0$ controls the dominant exponential decrease of the tail (iii) $r(m)>0$ for $m<1$, and $r(m)\to 0$ as $m \to 1$. 

Some of these numerical observations can be rationalized through the following analytical considerations. If one assumes the form (\ref{eq_tail_nu}), the exponential behavior of $\nu$ when $\tih \to +\infty$ (and also in $-\infty$ thanks to the symmetry of $\nu$) allows to continue $\hnu(z)$ from the real to the imaginary axis, i.e. to define the Laplace transform of $\nu$ as
\beq
L(t)=\hnu(-it) = \int \dd \nu(h) e^{t \tih} = \int \dd \nu(h) \cosh(t \tih) \ . 
\label{eq_Laplace}
\eeq
This function $L$ is well-defined for $t \in (-r(m),r(m))$, formally infinite for $|t| > r(m)$, with $L(0)=1$ and $L$ non-decreasing on $[0,r(m))$. Its behavior as $t \to r(m)$ depends on the exponent $\rho$ that controls the algebraic decay in (\ref{eq_tail_nu}). If $\rho \le 0$ then $L$ diverges in this limit, with a dominant behavior of $(r(m)-t)^\rho$ (and a prefactor that involves the slowly varying function $S$, see theorem XIII.5.1 in~\cite{Feller_vol2} for a rigorous statement of such a Tauberian argument), while if $\rho > 0$ the limit $L(r(m))$ is finite. We claim that for the probability measure $\nu$ solution of (\ref{eq_fp_nu}) one should have $\rho > 0$. Let us indeed rewrite this equation in terms of the Laplace transforms as $L(t)=\exp[R(t)]$, with $R(t) = \int \dd \pi(y) (e^{t y}-1) $. Consider now the tail behavior of the measure $\pi$ related to $\nu$ by (\ref{eq_fp_nu}); a simple computation, based in particular on the fact that $a(\tih)=\tih + O(e^{-2 \tih})$ as $\tih \to \infty$, shows that the asymptotic expansion (\ref{eq_tail_nu}) is also valid for the tail of $\pi$, up to the multiplicative coefficient $m$: $\pi(\{y \ge x \} ) \sim m \, \nu(\{\tih \ge x \} )$ as $x \to \infty$. As a consequence if $\rho < 0$ both $L(t)$ and $R(t)$ would diverge as $(r(m)-t)^\rho$ when $t \to r(m)$, which is clearly incompatible with the equation $L(t)=\exp[R(t)]$. We must thus have $\rho > 0$, in such a way that $L$ is finite in $r(m)$; nevertheless some derivatives of $L$ will be divergent in $r(m)$, each derivative with respect to $t$ in (\ref{eq_Laplace}) multiplying the integrand by $\tih$, hence reducing the exponent $\rho$ by 1. Suppose for simplicity that $0 < \rho \le 1$, in such a way that $L'(t)$ is divergent (higher values of $\rho$ would lead to the same conclusion by considering higher order derivatives). Because of the proportionality between the tails of $\nu$ and $\pi$ the function $R'(t)$ is also divergent, with $R'(t)/L'(t) \to m$ as $t \to r(m)$. Since $L'(t) = R'(t)\exp[R(t)] = R'(t) L(t) $, we are led to the conclusion that
\beq
1 = m \, L(r(m)) \ .
\label{eq_Tauberian}
\eeq
This type of reasoning can actually be put on rigorous grounds, the statement (\ref{eq_Tauberian}) being essentially the content of corollary 2.1 in~\cite{WaYa10}. As $L(r(m)) \ge L(0)=1$ we can conclude that a solution of (\ref{eq_fp_nu}) with a tail behavior of the form (\ref{eq_tail_nu}) can only exist for $m \le 1$, and that $r(m)$ must vanish when $m \to 1$, which confirms part of the numerical observations reported above. As $m$ is a decreasing function of $\g$ (see in particular Fig.~\ref{fig_mofgamma}) we are thus led by (\ref{eq_fp_m}) to the following definition for our conjectured value of $\gd$:
\beq
\gd = \int \dd \nu(h) \beta(h) \ ,
\eeq
where $\nu$ is the solution of (\ref{eq_fp_nu}) with $m=1$ (and hence has a power law decay when $\tih \to \infty$). A numerical evaluation of this quantity led us to the estimate $\gd \approx 0.871$. The property $m \to 1$ as $\g \to \gd$ was illustrated with the dotted horizontal line in the right panel of Fig.~\ref{fig_mn}, that corresponds to the expected height of the plateau of the curves in the critical region.

The square root behavior of $m$ as a function of $\g$ when $\g$ is close to $\gd$, that is clearly visible on Fig.~\ref{fig_mofgamma} and confirmed by a numerical fit of the data, is in line with the bifurcation of the functional equation (\ref{eq_fp_nu}). It is more surprising to observe that $\tC(\g)$ seems to behave linearly in the limit $\g \to \gd$ (see the right panel of Fig.~\ref{fig_largek}), as it could suggest that $\tC$ admits a continuation to smaller values of $\g$, in violation of the hypothesis made at the end of Sec.~\ref{sec_largen_numerical}. There are alternative explanations to this observation: the square root contribution could be very small and not observable with our numerical accuracy, or the expected square root behavior of $C(\alpha,k)$ when $\alpha \to \ad$ at finite $k$ could be washed out in the limit $k \to \infty$, i.e. the limits $\alpha \to \ad$ and $k\to \infty$ do not necessarily commute to correspond to the limit $\g \to \gd$.

\subsection{An analytic lowerbound on $\gd$}

As we have seen above our conjecture for $\gd$ is given in terms of the solution of a functional equation that does not seem to admit a simple expression. One can however derive explicitly a quantitatively close lowerbound, by observing that the function $\beta(h)$ defined in (\ref{eq_mnp1}) is such that $\beta(h) \ge 2 (\sqrt{2}-1)$ for all $h \in [-1,1]$. As a consequence the sequences of masses $m_n$ obey the inductive bounds $m_{n+1} \ge \exp[-\g + 2 (\sqrt{2}-1) m_n]$. The function $m \to \exp[-\g + 2 (\sqrt{2}-1) m]$ being increasing the sequence $\hm_n$ defined by $\hm_1 =e^{-\g}$, $\hm_{n+1} = \exp[-\g + 2 (\sqrt{2}-1) \hm_n]$ can be shown by induction to lowerbound the sequence $m_n$, namely $m_n \ge \hm_n$ for all $n$. It is then simple to show that for $\g < 1 + \ln (2 (\sqrt{2}-1)) \approx 0.812$ the sequence $\hm_n$ diverges with $n$, and as a consequence so does $m_n$. Under the assumption spelled out at the end of Sec.~\ref{sec_largen_numerical} on the equivalence of the divergence of $m_n$ and of $\tC_n$ this yields the lowerbound $\gd \ge 1 + \ln (2 (\sqrt{2}-1)) $.

\section{Conclusions and perspectives}
\label{sec_conclu}

We shall conclude this article by a brief summary and a list of possible directions for future works. We have computed the constant $\gd$ that appears in the asymptotic expansion for the clustering transition of two constraint satisfaction problems, the bicoloring of random $k$-hypergraphs in the limit $k \to \infty$ and the $q$-coloring of random graphs with $q \to \infty$ (we considered models with Poissonian degree distributions but our analysis would hold for any well-behaved distributions, in particular for regular graphs). Our computation relied on a series of assumptions, made explicit in Sec.~\ref{sec_reduced} and at the end of Sec.~\ref{sec_largen_numerical}, that were backed up with numerical simulations whenever possible. It would be of course highly desirable to obtain rigorous proofs of these hypotheses, thus enabling a quantitative enhancement of the results of~\cite{Sly08,MoReTe11_recclus,SlyZhang16}.

Another possible direction of work would be the study of other constraint satisfaction problems; the coincidence of $\gd$ for the two models studied here calls for an investigation of the range of universality of this constant. In particular the large $k$ limit of the reconstruction threshold of the random $k$-satisfiability problem~\cite{MontanariRicci08} would be worth investigating, because at variance with the two models considered here it does not have global symmetries like the spin flip invariance of the bicoloring problem (or the color permutation invariance for the $q$-coloring). In more technical terms its replica symmetric (RS) solution is non-trivial, this complication could thus affect the value of $\gd$ (unless the RS solution is sufficienly concentrated in the relevant regime of $\alpha$).

The computation we presented could also be generalized to the study of the 1RSB equations with Parisi breaking parameter distinct from 1, and/or to models at positive temperature. In the latter situation strictly hard fields would never exist, but the mechanism of soft fields that become asymptotically hard in the large $k$ limit of the regime $\alpha \in (\ad(k),\ar(k))$ could be at work also in this case. A related problem is the asymptotic expansion of the reconstruction threshold for the hardcore model on the tree~\cite{Brightwell_hardcore,BaKrZdZh13}, that has been studied in~\cite{Sly_hardcore} and presents a similar behavior.

Finally let us mention that our original motivation for the study presented here arose from our previous work~\cite{BuRiSe19} where we studied the possibility of tuning the value of $\ad(k)$ by considering non-uniform (biased) measures on the set of solutions of constraint satisfaction problems. For finite values of $k$ well-chosen biases allow indeed to increase $\ad$, thus improving the range of $\alpha$ in which solutions of random CSPs can be found by efficient algorithms. It remains to see whether this improvement survives the large $k$ limit, which requires being able to compute the asymptotics of $\ad(k)$ both in the unbiased and biased cases. The former was the object of the present article, work is currently under way to deal with the latter.

\acknowledgments

GS is part of the PAIL grant of the French Agence Nationale de la Recherche, ANR-17-CE23-0023-01.

\appendix

\section{The graph coloring case}
\label{sec_coloring}

This Appendix is devoted to the graph coloring problem. We shall present computations that are the counterparts of the ones explained in Sec.~\ref{sec_definitions} and \ref{sec_largek} of the main text for the hypergraph bicoloring problem, namely the definition of the reconstruction problem and its study in the limit where the number of colors $q$ and the degree $c$ diverge simultaneously according to (\ref{eq_scale}), and show that the very same equations summarized at the beginning of Sec.~\ref{sec_largen} arise in this limit. As the computations are quite similar we will be more succint than in the main text and concentrate on the specificities of the coloring problem.

Let us consider a rooted tree with spins $\s_i$ placed on its vertices. These spins can take $q$ values, interpreted as colors, $\s_i\in \{1,\dots,q \}$. A proper coloring of the tree is a configuration of the spins such that no edge is monochromatic (i.e. no pair of adjacent vertices are given the same color). A uniform proper coloring can be drawn in a broadcast fashion, by choosing the color $\s$ of the root uniformly at random among the $q$ possible ones, then each descendent of the root is assigned a color uniformly at random among the $q-1$ colors distinct from $\s$, and this is repeated recursively down to the $n$-th generation of the tree. In the reconstruction problem an observer is then provided with the colors on the vertices of the $n$-th generation of the tree only, and asked to guess the color of the root. The optimal strategy is to compute $\eta$, the posterior probability of the root given the observations, which is a distribution over $\{1,\dots,q\}$. A moment of thought reveals that the probability distribution of $\eta$, with respect to a broadcast process conditioned on the root value $\s$, and with respect to a random choice of the tree as a Galton-Watson branching process with Poisson offspring distribution of mean $c$, is a measure $P_{\s,n}(\eta)$ that can be determined recursively through the induction relation:
\beq
P_{\s,n+1}(\eta) = \sum_{l=0}^\infty e^{-c} \frac{c^l}{l!}
\frac{1}{(q-1)^l} \sum_{\s_1,\dots,\s_l \neq \s}
\int \prod_{i=1}^l \dd P_{\s_i,n}(\eta_i) \, \delta(\eta-f_{\rm c}(\eta_1,\dots,\eta_l))
\ ,
\label{eq_coloring_recursion}
\eeq
where the Belief Propagation recursion function $f_{\rm c}$ is defined here in such a way that $\eta=f_{\rm c}(\eta_1,\dots,\eta_l)$ means
\beq
\eta(\tau) = \frac{\underset{i=1}{\overset{l}{\prod}}(1-\eta_i(\tau))}{\underset{\tau'}{\sum} \underset{i=1}{\overset{l}{\prod}}(1-\eta_i(\tau'))} \ .
\label{eq_coloring_f}
\eeq
The initial condition of this recursive computation is given by $P_{\s,0}=\delta(\eta - \delta_\s)$, where $\delta_\s$ is the measure concentrated on the color $\s$, i.e. $\delta_\s(\tau) = \ind(\s=\tau)$, which corresponds to the colors being revealed on the leaves of the tree. These equations correspond to (\ref{eq_defs_fg},\ref{1RSBeq1_sigma},\ref{1RSBeq2_sigma}) for the hypergraph bicoloring problem.

We separate now the contribution of ``hard fields'', or frozen variables, namely the configurations of boundary variables that determine unambiguously the root in the naive reconstruction procedure. This forced value can only be the correct one the root had in the broadcast process, hence we shall write :
\beq
P_{\s,n}(\eta) = w_n \, \delta(\eta-\delta_\s) + (1-w_n) \, Q_{\s,n}(\eta) \ ,
\label{eq_coloring_def_hard}
\eeq
where $Q_{\s,n}$ has no atom on $\delta_\s$; this mimicks the decomposition (\ref{eq_bicol_def_hard}) of the main text. Plugging this decomposition in (\ref{eq_coloring_recursion}) yields the evolution equation for the weight of the hard fields:
\beq
w_{n+1} = \left(1-e^{-\frac{c w_n}{q-1}} \right)^{q-1} \ , \qquad \text{with} \ \ w_0=1 \ .
\label{eq_coloring_recursion_wn}
\eeq
Indeed $f_{\rm c}(\eta_1,\dots,\eta_l) = \delta_\s$ if and only if for each color $\s'\neq \s$ at least one of the arguments $\eta_i$ is equal to $\delta_{\s'}$, thus forbidding all colors except $\s$. The number of hard fields of the color $\s' \neq \s$ is easily seen from (\ref{eq_coloring_recursion}) to be Poisson distributed with average $\frac{c w_n}{q-1}$, independently from one color $\s'$ to another, from which (\ref{eq_coloring_recursion_wn}) follows.

The recursion (\ref{eq_coloring_recursion_wn}) has a bifurcation at $c_{\rm r}(q)$, in the sense that $w_n \to 0$ as $n \to \infty$ if and only if $c<c_{\rm r}(q)$. Writing down the equations fixing $c_{\rm r}$ and $w_{\rm r}$ at the bifurcation, which are similar to (\ref{eq_wr_Gammar}), and then expanding them for large $q$ one finds the asymptotic expansion $c_{\rm r}=q (\ln q + \ln \ln q + 1 + o(1))$. We shall thus study the large $q$ limit with $c$ getting also large, on the scale $c=q (\ln q + \ln \ln q + \g)$ with $\g$ finite. In this limit one finds that for $n$ finite 
\beq
w_n = 1 - \frac{x_n}{\ln q} + o\left(\frac{1}{\ln q} \right) \ ,
\label{eq_coloring_scaling_wn}
\eeq
where $x_n$ is of order 1 and obeys exactly the same recursion as in the bicoloring case, namely $x_0=0$ and $x_{n+1}=e^{-\g + x_n}$.

Let us now simplify the evolution equation for the distributions $Q_{\s,n}$ of the soft fields. First of all we insert the decomposition (\ref{eq_coloring_def_hard}) in the right hand side of (\ref{eq_coloring_recursion}) and obtain, without any approximation,
\begin{align}
P_{\s,n+1}(\eta) =& 
\sum_{l=0}^\infty e^{-c(1-w_n)} \frac{(c(1-w_n))^l}{l!} \frac{1}{(q-1)^l} \sum_{\s_1,\dots,\s_l \neq \s} \sum_{\{p_{\s'}=0,1\}_{\s' \neq \s}} \prod_{\s' \neq \s} \left(e^{-\frac{c w_n}{q-1}}\right)^{p_{\s'}} \left(1-e^{-\frac{c w_n}{q-1}}\right)^{1-p_{\s'}} \nonumber \\
& 
 \int \prod_{i=1}^l \dd Q_{\s_i,n}(\eta_i) \, \delta(\eta-\widetilde{f}_{\rm c}(\s,\{p_{\s'}\}_{\s' \neq \s};\eta_1,\dots,\eta_l)) \ ,
\label{eq_coloring_decomposed}
\end{align}
where $l$ is the number of neighbors of the root that receive a soft field, $\s_1,\dots,\s_l$ the colors these vertices have in the broadcast, and the indicator variables $p_{\s'}$ are equal to 1 if and only if no neighbor of the root assigned the color $\s'$ in the broadcast is perfectly recovered (i.e. receives a hard field). Hence the colors $\s'\neq \s$ with $p_{\s'}=0$ are precisely the ones forbidden for the root, as at least one of its neighbors is forced to this value. The relation $\eta=\widetilde{f}_{\rm c}(\s,\{p_{\s'}\}_{\s' \neq \s};\eta_1,\dots,\eta_l)$ is obtained by specializing $f_{\rm c}$ of (\ref{eq_coloring_f}) to this pattern for the presence of hard fields in its arguments, and thus reads
\beq
\eta(\tau) = \frac{p_\tau\underset{i=1}{\overset{l}{\prod}} (1-\eta_i(\tau))}{\underset{\tau'}{\sum} p_{\tau'}\underset{i=1}{\overset{l}{\prod}}(1-\eta_i(\tau'))} \ ,
\eeq
with the convention $p_{\s}=1$.

Let us call $p =\underset{\s' \neq \s}{\sum} p_{\s'}$ the number of colors that satisfy the condition explained above; in the equation (\ref{eq_coloring_decomposed}) it corresponds to a random variable with a binomial distribution of parameters $(q-1,e^{-\frac{c w_n}{q-1}})$. According to (\ref{eq_coloring_scaling_wn}) the product of these parameters go to zero as $1/\ln q$ in the limit we are considering, we shall thus truncate (\ref{eq_coloring_decomposed}) on the smallest possible values of $p$.
As $p=0$ yields a hard field in the left hand side of (\ref{eq_coloring_decomposed}), the distribution of the soft fields is dominated in this limit by the case $p=1$. As a consequence the fields $\eta$ in the support of $Q_{\s,n}$ have non-zero values on two colors only, $\s$ and another one $\s'$ uniformly distributed on the $q-1$ possibilities. Let us parametrize this type of distributions via a distribution $Q_n(h)$ on real random variables $h\in[-1,1]$, 
\beq
Q_{\s,n}(\eta) = \int \dd Q_n(h) \frac{1}{q-1} \sum_{\s' \neq \s} \delta(\eta-s(\s,\s';h)) \ , \qquad \text{with} \quad
s(\s,\s';h)(\tau) = \begin{cases}
  \frac{1+h}{2} & \text{if} \ \ \tau = \s \\
\frac{1-h}{2} & \text{if} \ \ \tau = \s' \\
0 & \text{otherwise}
\end{cases} \ .
\label{eq_coloring_parametrization}
\eeq
Let us also denote $\widehat{f}_{\rm c}(\s,\s';\eta_1,\dots,\eta_l)$ the function 
$\widetilde{f}_{\rm c}(\s,\{p_{\s''}\}_{\s'' \neq \s};\eta_1,\dots,\eta_l)$ with $p_{\s''}=\delta_{\s'',\s'}$, in such a way that the only two non-zero components of $\widehat{f}_{\rm c}(\s,\s';\eta_1,\dots,\eta_l)$ correspond to the colors $\s$ and $\s'$. Using this notation, and the parametrization (\ref{eq_coloring_parametrization}), one can deduce from (\ref{eq_coloring_decomposed}) the evolution equation for $Q_{\s,n}(\eta)$ at lowest order:
\begin{align}
Q_{\s,n+1}(\eta) =& \frac{1}{q-1}\sum_{\s' \neq \s}
\sum_{l=0}^\infty e^{-c(1-w_n)} \frac{(c(1-w_n))^l}{l!} 
 \prod_{i=1}^l \left( \frac{1}{(q-1)^2} \sum_{\substack{\s_i\neq \s \\ \s'_i \neq \s_i}} \int \dd Q_n(h_i) \right) \label{eq_coloring_simplified} \\ & \delta(\eta-\widehat{f}_{\rm c}(\s,\s';s(\s_1,\s'_1;h_1),\dots,s(\s_l,\s'_l;h_l))) \nonumber
\end{align}
To put this expression under the form (\ref{eq_coloring_parametrization}), and hence close the recursion on $Q_n(h)$, it remains to notice that
\beq
\widehat{f}_{\rm c}(\s,\s';\eta_1,\dots,\eta_l) = s(\s,\s',h) \quad \text{with} \ \ h=f(u_1,\dots,u_l) \ , \quad  u_i = \widehat{u}(\s,\s';\eta_i) \ ,
\eeq
where $f$ is the function defined for Ising spins in (\ref{eq_defs_fg}), and
\beq
\widehat{u}(\s,\s';\eta) = \frac{\eta(\s')-\eta(\s)}{2-\eta(\s)-\eta(\s')} \ .
\eeq
For a fixed choice of $\s$ and $\s' \neq \s$, one can see that $\widehat{u}(\s,\s';s(\s_i,\s'_i;h_i))$ is a random variable with respect to the uniform choices of $\s_i \neq \s$ and $\s'_i \neq \s_i$, that takes the following values (recall the definition of the function $g_1$ from (\ref{eq_def_g1})):
\beq
\begin{cases}
  g_1(h_i) & \text{with probability} \ \frac{q-2}{(q-1)^2} \ , \text{when} \ \s_i =\s' \ , \ \s'_i \neq \s \ , \\
  g_1(-h_i) & \text{with probability} \ \frac{q-2}{(q-1)^2} \ , \text{when} \ \s'_i =\s' \ , \ \s_i \neq \s \ , \\
  -g_1(-h_i) & \text{with probability} \ \frac{q-2}{(q-1)^2} \ , \text{when} \ \s'_i =\s \ , \ \s_i \neq \s' \ , \\
  h_i & \text{with probability} \ \frac{1}{(q-1)^2} \ , \text{when} \ \s'_i =\s \ , \ \s_i = \s' \ , \\
  0 & \text{otherwise} \ .
\end{cases}
\eeq
As $c(1-w_n)\frac{q-2}{(q-1)^2} \to x_n $ in the regime we are considering, the number of occurences of the first three cases in (\ref{eq_coloring_simplified}) is Poissonian of mean $x_n$; on the other hand the number of times the fourth case happens vanishes when $q$ diverges (as $1/q$), while the fifth does not contribute to (\ref{eq_coloring_simplified}), because $f(u_1,\dots,u_l,0,\dots,0)=f(u_1,\dots,u_l)$. We thus see by comparison with (\ref{eq_summary_Qn},\ref{eq_summary_hRn}) that the probability distribution $Q_n(h)$ defined in (\ref{eq_coloring_parametrization}) obeys exactly the same recursion equations as the one derived in the main text for the hypergraph bicoloring model, the initial condition $Q_1(h)=\delta(h)$ being also valid here.

\bibliography{biblio}

\end{document}